\documentclass[prb,showpacs,longbibliography,notitlepage,floatfix]{revtex4-1}
\usepackage{amsmath}
\usepackage{amsfonts}
\usepackage{stmaryrd}
\usepackage{bbm}
\usepackage{mathrsfs}
\usepackage{tipa}
\usepackage{amssymb}
\usepackage{txfonts}
\usepackage{graphicx}
\usepackage{dcolumn}
\usepackage{epstopdf}
\usepackage[colorlinks,linkcolor=blue,urlcolor=blue,citecolor=blue]{hyperref}
\usepackage{multirow}
\usepackage{subfigure}
\usepackage{color}
\usepackage{graphicx}

\begin{document}
\newcommand*{\cm}{cm$^{-1}$\,}
\newcommand*{\Tc}{T$_c$\,}

\title{Infrared properties of heavy-fermions: evolution from weak to strong hybridizations}

\author{R. Y. Chen}
\author{N. L. Wang}
\affiliation{International Center for Quantum Materials, School of Physics, Peking University, Beijing 100871, China}
\affiliation{Collaborative Innovation Center of Quantum Matter, Beijing 100871, China}

\begin{abstract}
In this article, we review the charge excitations of heavy fermion compounds probed by infrared spectroscopy. The article is not meant to be a comprehensive survey of experimental investigations. Rather it focuses on the dependence of charge excitations on the hybridization strength. In this context, the infrared properties of the Ce$_m$M$_n$In$_{3m+2n}$ family are discussed in detail since the hybridization strengths differ dramatically in different members despite their similar lattice structures. Investigations on some mixed valent compounds are also presented aiming to elucidate the generic trend on the evolution. In particular, we address the scaling between hybridization energy gap $\Delta_{dir}$ and the hybridization strength $\tilde{V}$($\propto\sqrt{WT_K}$) in a wide range of heavy fermions compounds which demonstrates that the periodic Anderson model can generally and quantitatively describe the low-energy charge excitations.

\end{abstract}


\maketitle

\section{Introduction}
Optical spectroscopy is a primary technique to investigate charge dynamical properties and band structure of materials as it probes both free carriers and interband excitations. In particular, it yields direct information about formation of energy gaps. The electrodynamic response of heavy fermion (HF) metals has been studied in a large number of systems, and many of them were discussed in several review articles \cite{Degiorgi1999,Riseborough2000,Basov2011}. In the present article, we do not intend to present a thorough review on the existing investigations, instead we focus on the evolution of charge excitation properties as a variation of hybridization strength. Our attention is limited to heavy fermion or mixed valent metals. Therefore, the Kondo semiconductors \cite{Riseborough2000}, which is a special case that the Fermi level locates coincidentally within the hybridization gap, are not included in the present discussions. For our purpose here, the Ce$_m$M$_n$In$_{3m+2n}$ family offers a particularly good opportunity for exploring such evolution from weak to strong hybridizations since different members in the family exhibit different ground state properties. We shall also present optical investigations on some mixed valent compounds which help to elucidate the generic trend on the evolution. Before we review those works, we explain some fundamental physics related to the heavy fermions in the introduction. Those include the Kondo versus RKKY interations, the Doniach phase diagram and the periodical Anderson model, which should help readers to understand the related spectroscopic features.

\subsection{Kondo versus RKKY interactions}
Heavy fermions compounds contain a lattice of strongly correlated $f$-electrons and a sea of weakly interacting conduction
electrons. The $f$-electrons are associated with the rare-earth or actinide ions in the lattice and are localized due to the absence of direct overlap between their wave functions. They carry magnetic moments and give rise to a susceptibility that is of the Curie-Weiss form.
On the other hand, the extended conduction electrons interact with the local moments of $f$-electrons, being referred to as Kondo interaction.

For isolated magnetic impurities, Kondo interaction causes a spin-flip scattering of conduction electron at high temperature, which explains the logarithmic temperature dependence of resistivity in dilute magnetic alloys.\cite{Gruner1974} But at low temperature, this interaction causes the conduction electrons to collectively screen the local magnetic moment of \emph{f}-electron to form a Kondo singlet. Its effect is to build a narrow resonance feature, with a width of Kondo temperature ($T_K$) , in the density of states (DOS) at the Fermi level, often referred to as Abrikosov-Suhl or Kondo resonance\cite{Abrikosov1965,Suhl1965,Johansson1978,Peterman1983,Moshchalkov1985,Coleman2006,Coleman2015}. The Kondo temperature, being associated with the screening effect, has the order of $k_BT_K\propto W exp[-1/{JD(E_F)}]$ , where the \emph{J} is Kondo coupling strength, $W$ the conduction electron bandwidth, and $D(E_F)$ the density of states near Fermi level\cite{Doniach1977}.

\begin{figure}
  \centering
  \includegraphics[width=8.2cm]{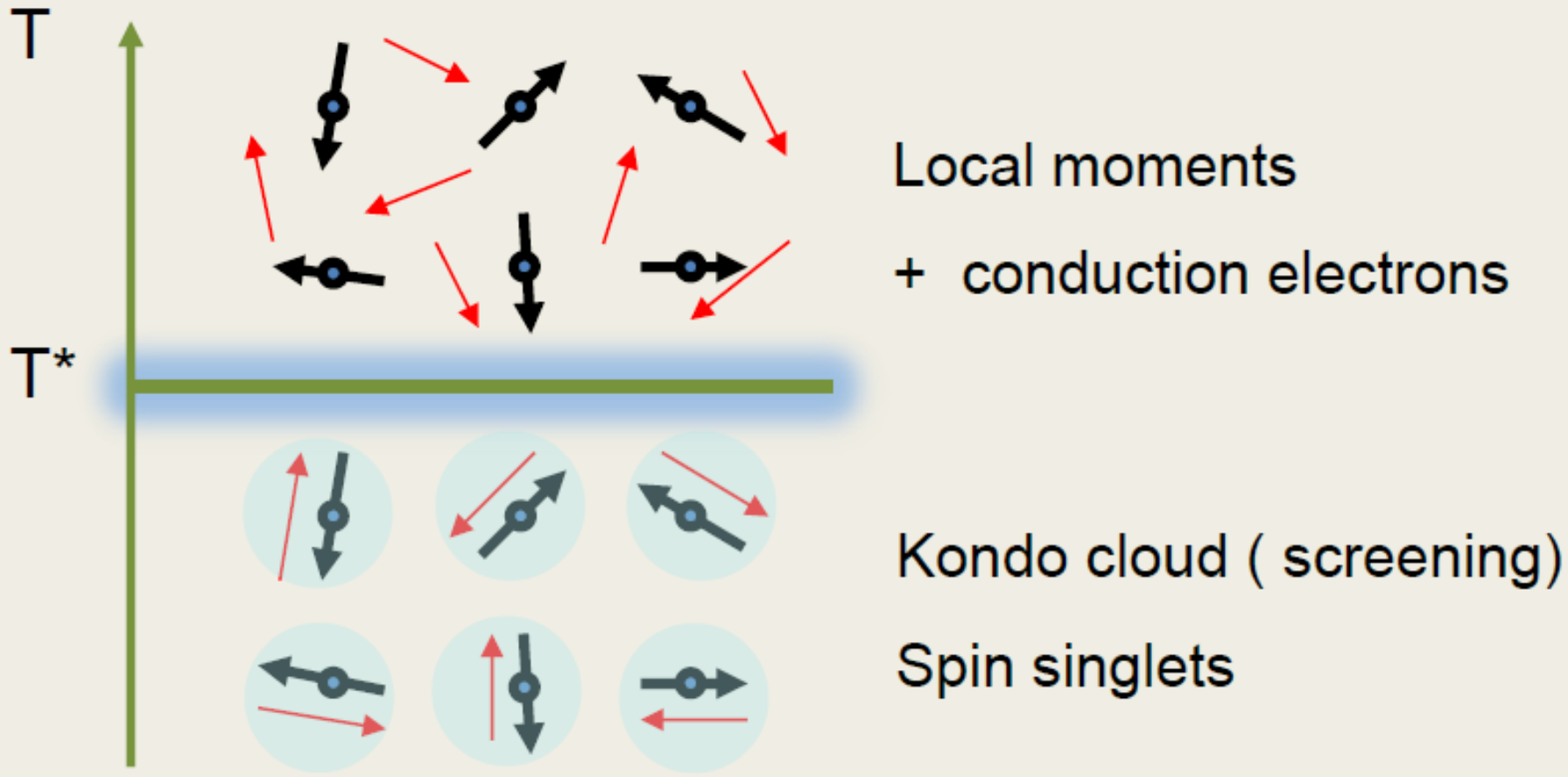}\\
  \caption{Kondo screening formation in a heavy fermion metal. Balck thick arrow refers to the local moment formed by f-electrons, and red thin arrow refers to the spin state of a conduction electron.}\label{Fig:Kondoscreening}
\end{figure}

In heavy fermion systems, where the local moments are periodically arranged, the Kondo resonance at each $f$-electron site transforms into a narrow electron band. This narrow band hybridizes with the conduction electron band, leading to the formation of a hybridization energy gap. Since the local magnetic moments of $f$-electrons evolve into itinerant behavior, the $f$-electrons should be counted in
the Fermi volume at low temperature, and the corresponding Fermi surface becomes ¡°large¡± as a consequence. Figure \ref{Fig:Kondoscreening} shows a schematic picture of the Kondo screening formation in a heavy fermion metal at low temperature. The screening effect leads to a paramagnetic and constant spin susceptibility.

\begin{figure}
  \centering
  \includegraphics[width=8cm]{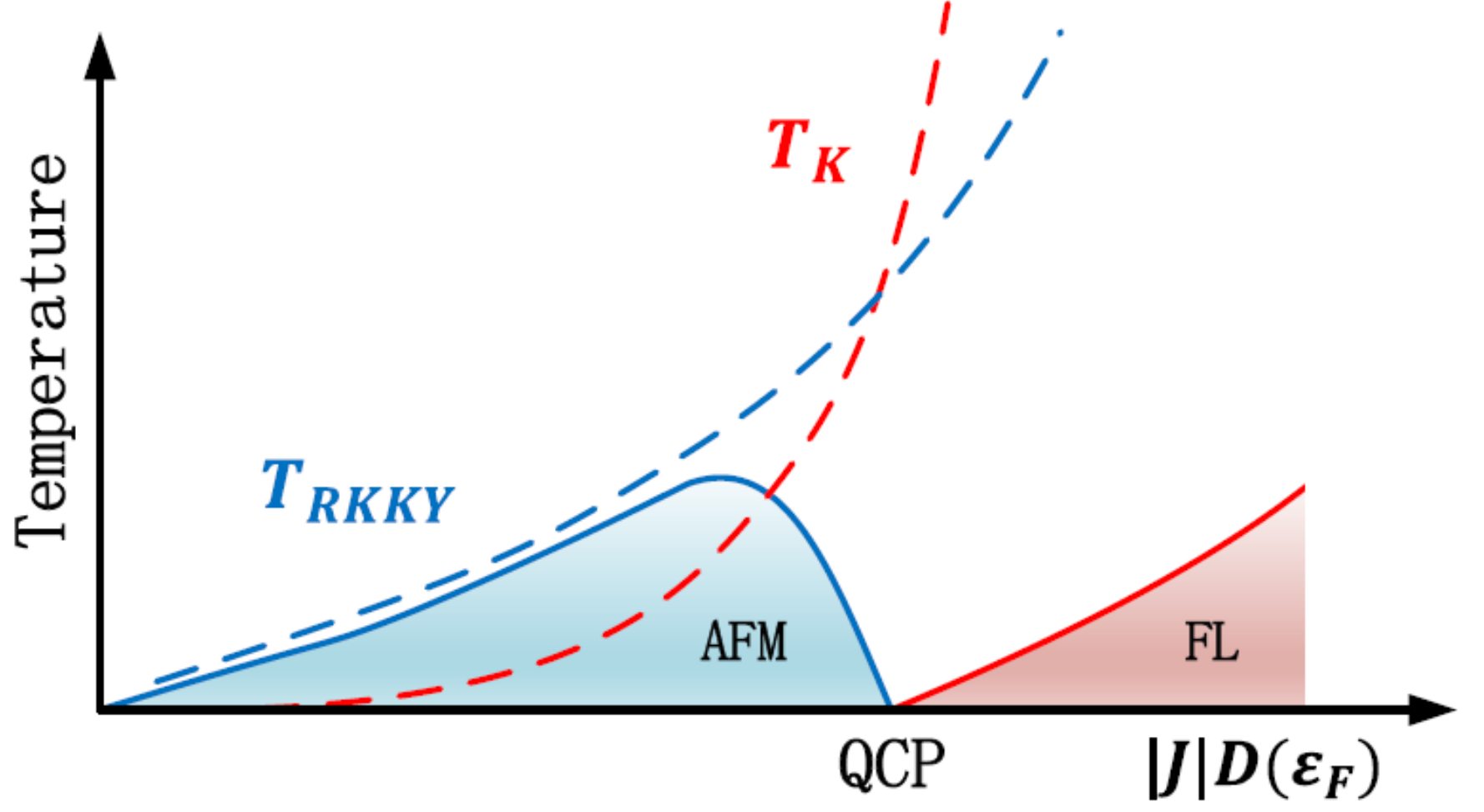}\\
  \caption{Doniach phase diagram for the Kondo lattice, illustrating the antiferromagnetic regime and the heavy fermion liquid regime for $T_K<T_{RKKY}$ and $T_K>T_{RKKY}$, respectively.}\label{Fig:Doniach}
\end{figure}

In addition to the Kondo interaction, there also exists the Ruderman-Kittel-Kasuya-Yoshida (RKKY) interaction between the local moments, which is mediated by the conduction electrons, that tends to order the local moments \cite{Ruderman1954,Kasuya1956,Yoshida1957}. Its strength can be expressed by the characteristic temperature $T_{RKKY}$ which satisfies the relation $k_BT_{RKKY}\propto J^2D(E_F)$. The competition between the demagnetizing on-site Kondo interaction and the intersite RKKY exchange interaction is believed to be the key factor that determines the ground state properties. For small values of $J$, the relation $T_{RKKY}>T_K$ holds because of the exponential dependence of $T_K$  on $J$. In this situation, magnetic order is more favorable. For large values of $J$ the energy gained by the formation of local Kondo singlets will surpass the gain due to the magnetic ordering. Consequently, a nomagnetic ground state without long-range magnetic order will develop, although short-range order or magnetic spin flluctuations may exist. Close to the critical value $J_c$, where the two characteristic temperature $T_{RKKY}$  and $T_K$ are equal, non-Fermi liquid properties are often observed\cite{Tsvelik1993,Lohneysen1994,Lohneysen1995,Lohneysen1996,Steglich1996,Nakotte1996,Buttgen1996,Steglich1997,Pietri1997,Gegenwart1999,Trovarelli2000}. Such a competition is generally described by the so-called Doniach phase diagram\cite{Doniach1977}, as is shown in Fig. \ref{Fig:Doniach}.  It deserves to remark that the scale of the diagram depends critically on the magnitude of the spin. For spin 1/2, the quantum critical point (QCP) would occur for unrealistically large values of J (J=W). For a more physical phase diagram one needs to consider larger spin degeneracy. Coleman showed that for rare-earth f-electron based compounds, the large spin degeneracy N=2j +1 (where j is the total angular momentum arising from orbital and spin coupling) would greatly enhance the local spin fluctuation and, as a result, stabilize the Kondo lattice ground state with a much smaller Kondo coupling constant \cite{Coleman1983}.

Since the physics of heavy fermion compounds is determined by a delicate balance and competition between the RKKY-type magnetic interaction and the Kondo screening interaction (\emph{i.e.} hybridization of the 4\emph{f} or 5\emph{f} electrons with delocalized conduction states), it is useful to look into the evolution of physical properties as a function of hybridization strength or compounds with different ground states. According to this point of view, the heavy fermion compounds could be roughly divided into three categories: the heavy fermion antiferromagnets, the paramagnetic heavy fermion metals, and mixed valent compounds. Their coupling strengths increase in sequence. We will discuss the optical responses for each of them, by taking a few examples in the corresponding category.

\subsection{Periodic Anderson model and hybridization energy gap}
Theoretical treatment of heavy fermions or intermediate valence compounds centers on the periodic Anderson model (PAM), in which the hybridization between correlated \emph{f}-electrons and non-interacting conduction band is considered\cite{Hewson1993,Newns1987,Coleman2006,Logan2005}. In the strong coupling limit, the model is reduced to the Kondo lattice model. The PAM captures most of the spectral features in optical conductivity for heavy fermions. Before we review the optical measurement results on the evolution of hybridization strength, we shall first briefly introduce the key results related to optical response.

\begin{figure}
  \centering
  \includegraphics[width=8.5cm]{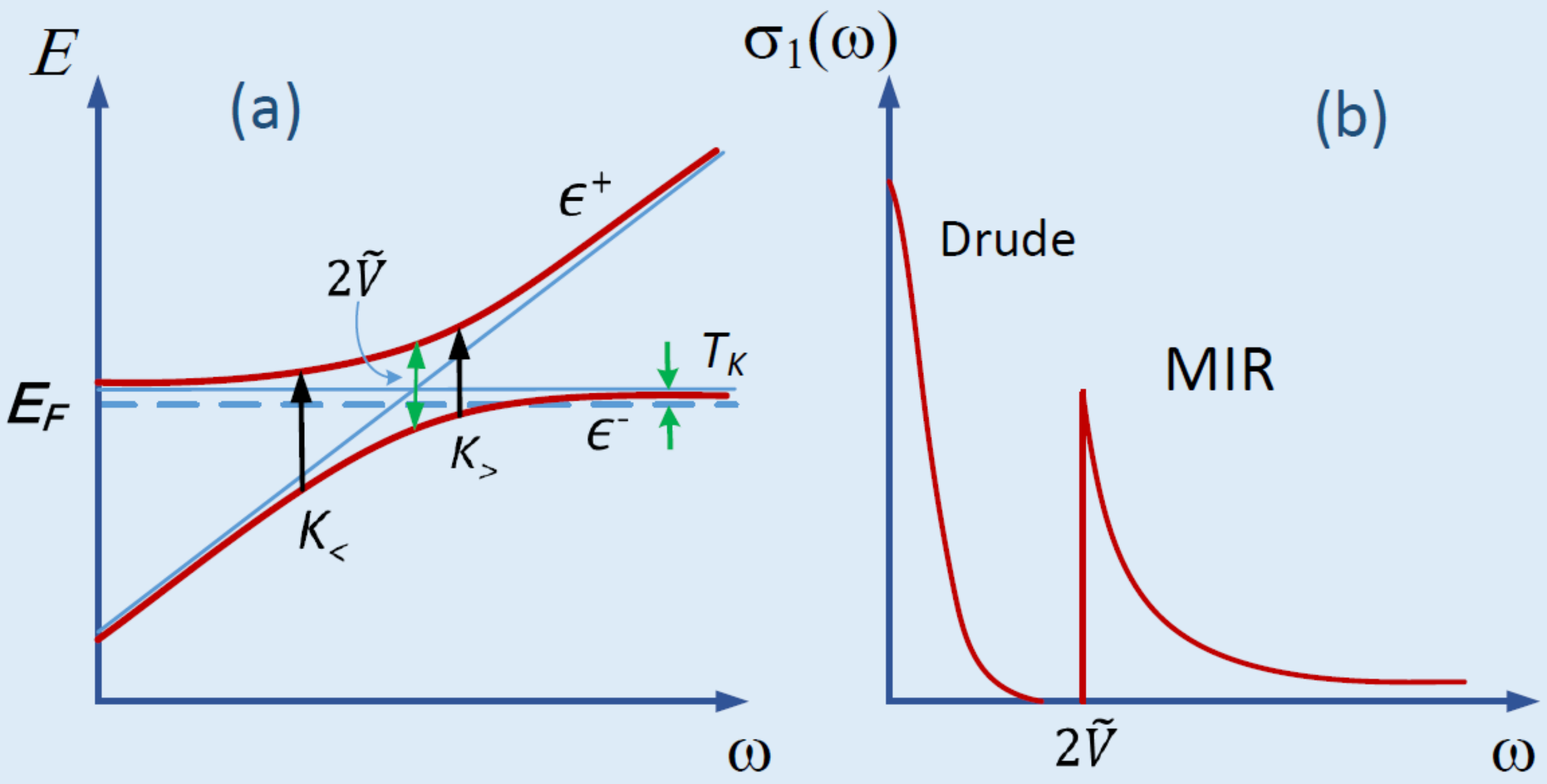}\\
  \caption{Schematic diagram for the mean field approximation results of PAM: (a) dispersions of the renormalized hybridized bands near $E_F$ with arrows indicating transitions allowed in the existence of the Kondo resonance; (b) the real part of optical conductivity expected from the two hybridized bands $\epsilon^+$ and $\epsilon^-$. Broadening from the finite lifetime effect of quasiparticles is not considered.}\label{Fig:PAL}
\end{figure}

The Hamiltonian for the PAM is given by\cite{Hewson1993}
\begin{equation}
\begin{split}
H=  \sum_{i,\sigma}\epsilon_k c_{k,\sigma}^{\dag} c_{k,\sigma} +\sum_{i,\sigma}\epsilon_f f_{i,\sigma}^{\dag} f_{i,\sigma}+U\sum_i n_{f,i,\downarrow} n_{f,i,\uparrow} \\
 + \sum_{i,k,\sigma}(V_k e^{i k\cdot R_i} f_{i,\sigma}^{\dag} c_{k,\sigma} +H. C).
\label{PAM1}
\end{split}
\end{equation}
The first term represents the uncorrelated conduction electron band, the second and the third term describe the correlated $f$-levels, and the final term represents the hybridization between the conduction band and $f$-levels. The PAM cannot be solved exactly. A mean-field approximation to the model captures its essential physics which we shall briefly introduce below. Other approximate solutions of PAM have been reported, including explicit
calculations of the $T$-dependent optical conductivity based on the dynamical mean field theory \cite{Logan2005}, and the physical results obtained are qualitatively the same. It is well known that the application of mean-field approximation to PAM leads to two renormalized bands \cite{Hewson1993,Newns1987},
\begin{equation}
\epsilon^{\pm}={[ \epsilon_k + E_F +\tilde{\epsilon}_f  \pm \sqrt{(E_F + \tilde{\epsilon}_f - \epsilon_k )^2 + 4\tilde{V}^2}]}/2
\label{PAM2}
\end{equation}
Here, the $\tilde{\epsilon}_f $ is related to the renormalized \emph{f} level and $\tilde{V}$=$z^{1/2}V$ is the
hybridization strength renormalized by the on-site $f$-electron repulsion with $z=1-n_f$ ($n_f$, $f$-level occupancy). This energy scales with the Kondo temperature as $\tilde{V}\sim\sqrt{T_k W}$, where $W$ is the conduction electron bandwidth \cite{Hewson1993,Newns1987,Coleman2006,Logan2005}. As is shown in Fig.\ref{Fig:PAL}(a), a hybridization gap opens in \textbf{k}-space with a direct gap of
\begin{equation}
\Delta_{dir}=2\tilde{V} \propto \sqrt{T_k W},
\label{PAM3}
\end{equation}
Meanwhile, an indirect gap would appear in the density of states. It has much smaller energy scale and is given as $\Delta_{ind}$=2$\tilde{V}^2/W\propto T_K$. The optical response detects the direct hybridization energy gap.

The optical conductivity could be calculated from the Kubo-Greenwood formula. In the case that there are two relevant bands, $\epsilon^+$ and $\epsilon^-$, and the real part of optical conductivity could be rewritten as\cite{Hancock2004},
\begin{equation}
\sigma_{pam}={{e^2} \over{4\pi^2m^2\omega}}\int_{\Delta \epsilon=\omega}{dS \over{\mid \nabla_k(\epsilon^+ -\epsilon^-) \mid}}\mid \textbf{P}_{+,-} \mid^2
\label{PAM4}
\end{equation}
where $\textbf{P}_{+,-}$ is the dipole matrix element connecting the two bands $\epsilon^+$ and $\epsilon^-$. For each value of $\omega >2\tilde{V}$, there are two contributions to the integral: one originating from the levels inside the unrenormalized Fermi surface ($k_<$); the other from the
states occupied as a result of renormalization, i.e., the heavy flat band ($k_>$).  Figure \ref{Fig:PAL}(b) shows a sketch of real part of optical conductivity based on PAM approach. The extremely sharpness across the direct energy gap is due to the infinite lifetime for the quasiparticle states. In reality the finite lifetime of quasiparticles should smear the sharpness and results in a broad peak.

 In general, optical spectroscopy measurements on heavy electron systems should contain two features: a very narrow Drude peak at low frequency contributed from heavy quasiparticles crossing the Fermi level E$_F$ and excitations across a direct hybridization gap ($\Delta_{dir}$).  The spectral weight of Drude component is proportional to $\omega_{pHF}=4\pi n_{HF} e^2/m^*$. The development of a hybridization energy would lead to a suppression of spectral weight below the gap energy scale and a peak above the gap energy scale, commonly referred to as mid-infrared peak. The mean-field approximation to the PAM also predicts that the effective mass of the heavy electrons should scale as square of the ratio between the hybridization gap and the characteristic scale $T_K$ of the heavy Fermi liquid
\begin{equation}\label{Eq:PAM5}
  m^*/m_e = ( \Delta_{dir}/T_K)^2.
\end{equation}
This relation was verified by optical study on a number of HF compounds\cite{Dordevic2001}.

The hybridization gap is one of the most important characteristic features in heavy fermion systems, which is linked to the hybridization strength according to PAM. In order to study the evolution from weak to strong hybridizations in different materials and explore the underlying mechanism, it is effective to summarize and compare their optical responses, which directly reflect the energy scale and resonance strength of hybridization gaps. Specifically, infrared spectroscopy is also quite sensitive to the variation of free carriers, yielding information about the plasma frequency of itinerant quasiparticles, a quantity linked to the ratio of carrier number over carrier effective mass. In the main text hereinbelow, we will first demonstrate the infrared optical properties of the heavy fermion Ce$_m$M$_n$In$_{3m+2n}$ family, which includes both antiferromagnets (CeIn$_3$, CeRhIn$_5$ and CePt$_2$In$_7$) and paramagnetic metals (CeCoIn$_5$ and CeIrIn$_5$). Then we will turn to some mixed valent examples: such as YbAl$_3$, YbIn$_{1-x}$Ag$_x$Cu$_4$, EuIr$_2$Si$_2$ and EuNi$_2$P$_2$. Significantly, one can find the presence of an universal scaling property between the hybridization energy gap and the hybridization strength for Ce- and Yb-based heavy fermions.

Before presenting experimental data, we would like to remark that Equ.\ref{PAM1} should be identified as the spin half version of the Anderson Lattice Model. The spin degeneracy of f-electrons is not taken into account here. Although the characteristic features in optics are captured by the mean-field solution of the model, the f-electron spin degeneracy indeed has some effects on the experimental observations in optics. On the one hand, the large N=2j+1 spin degeneracy would modify the expression of Kondo temperature, an effect of which will be discussed in Section III (c); on the other hand, the crystal field can act on the spin degeneracy, resulting in a change of hybridization strength. Such effect will be discussed in the next session.

\section{Infrared properties of Ce$_m$M$_n$In$_{3m+2n}$}
\label{sec:IntroCMI}

 \begin{figure}[htbp]
  \centering
  \includegraphics[width=7cm]{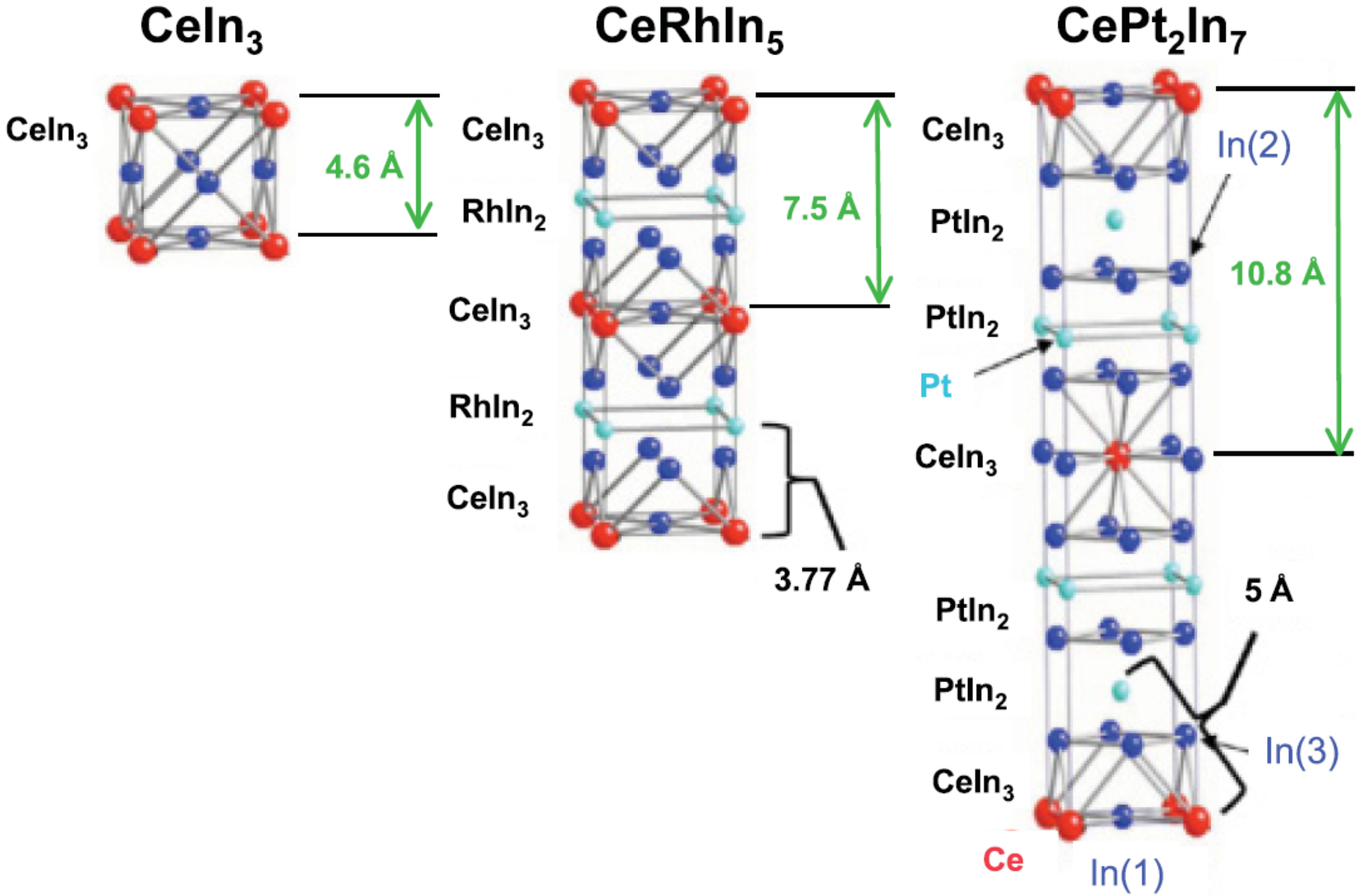}\\
  \caption{Schematic crystal structures of CeIn$_3$, CeRhIn$_5$, and CePt$_2$In$_7$ with interlayer spacings indicated.\cite{Altarawneh2011}}\label{Fig:altarawneh1}
\end{figure}

Among the most-studied heavy fermion compounds are the Ce-based 115 CeMIn$_5$ (M=Co, Rh, Ir). They belong to a family Ce$_m$M$_n$In$_{3m+2n}$ where m is the number of CeIn$_3$ layers, n is the number of MIn$_2$ layers and M is a transition
metal (Co, Rh, Ir, Pd or Pt). They may be regarded structurally as an alternate stacking of CeIn$_3$ and MIn$_2$ building blocks along the c-axis. These Ce-based compounds are close to antiferromagnetic quantum critical points and exhibit rich phase diagrams. In the 115 compounds (m=1, n=1), CeCoIn$_5$ and CeIrIn$_5$ are heavy fermion metals and superconduct at ambient pressure. By contrast, CeRhIn$_5$ shows antiferromagnetism below $T_N=$3.8 K, though some studies revealed that CeRhIn$_5$ also becomes superconducting below 0.1 K, deep within the antiferromagnetic (AFM) phase \cite{Paglione2008}. On the other hand, both CeIn$_3$ (m=1, n=0) and CePt$_2$In$_7$ (m=1, n=2) have AFM ground states with Neel temperatures of $T_N=$10 K and 5.5 K, respectively. The different ground states reflect different ratios of the hybridization to RKKY interactions. Therefore, the Ce$_m$M$_n$In$_{3m+2n}$ family offers a good opportunity to study the evolution of the charge excitation properties from weak to strong hybridizations. Figure \ref{Fig:altarawneh1} shows the crystal structures of three representative members in the family.

\subsection{Infrared properties of CeMIn$_5$ M=Co, Rh, Ir}

For a conventional heavy fermion metal, which locates at the right side of the Doniach phase diagram, the ground state belongs to a traditional Fermi liquid state. The resistivity of this kind of materials at very low temperatures follows quantitatively the $T^2$ law, although a maximum is regularly observed at higher temperatures, because the the Kondo screening effect substantially reduces the scattering rate raised by magnetic moments. The specific heat of such intermetallic compound usually yields a very large sommerfeld coefficient $\gamma$, in associated with the extremely "heavy" quasiparticles. Of significant importance, many heavy fermion metals exhibit unconventional superconductivity at very low temperatures, the initial discovery of which in CeCu$_2$Si$_2$\cite{Steglich1979} had definitely opened up a new frontier in the study of strongly correlated electron systems.

\begin{figure}
  \centering
  \includegraphics[width=7cm]{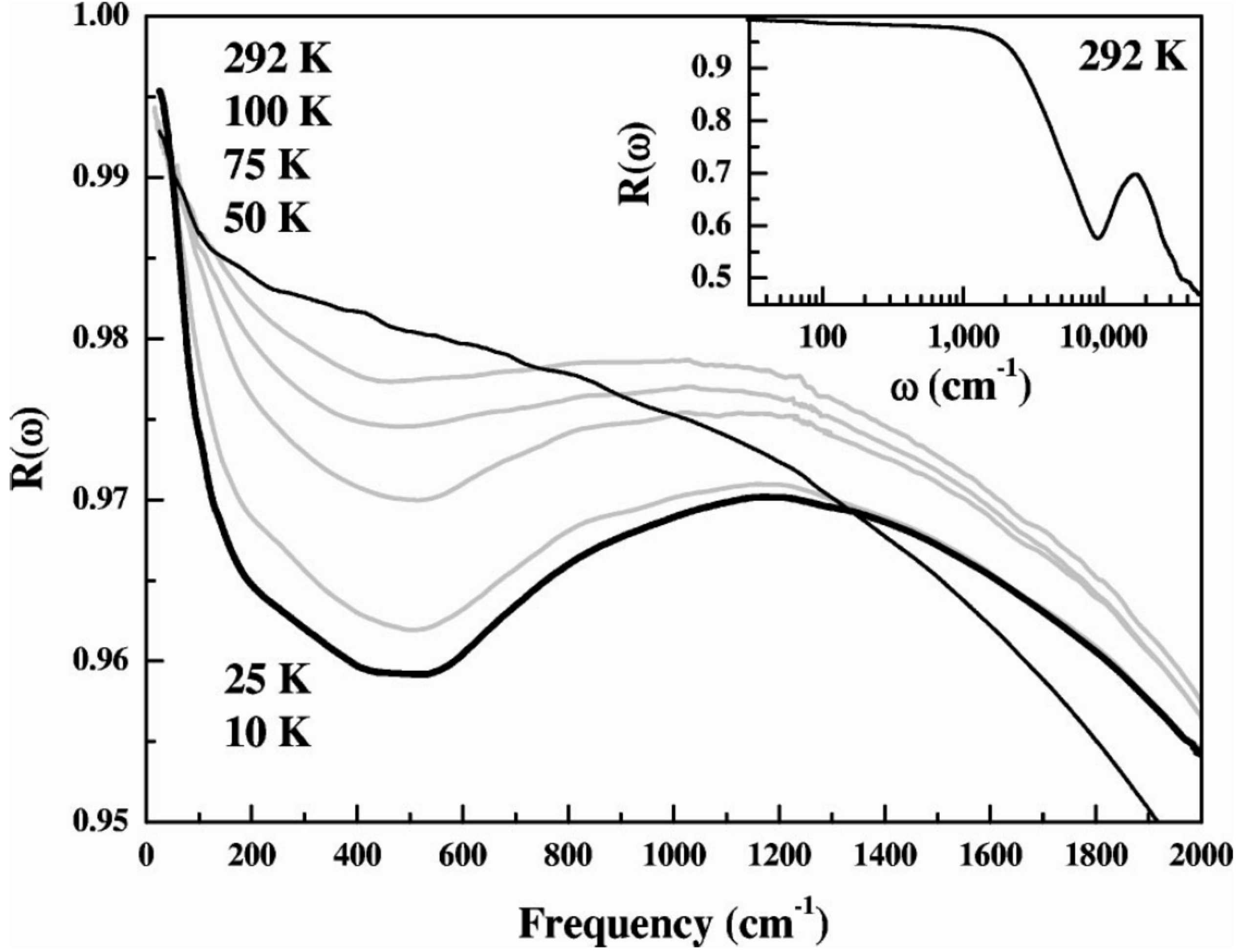}\\
  \caption{Temperature dependence of the reflectance in the far infrared. At 292 K, R($\omega$) increases smoothly showing metallic behavior. At lower temperatures a minimum develops in R($\omega$) at
500 \cm. The inset shows the reflectance at 292 K over the entire
measured frequency range on a log scale.\cite{Singley2002}}\label{Fig:singley1}
\end{figure}

Let us first examine the optical measurement results on CeCoIn$_5$ which is a heavy fermion metal and becomes superconducting below 2.3 K \cite{Petrovic2001,Nicklas2001}. The infrared properties of CeCoIn$_5$ was first reported by Singley et al. within the temperature range of 10 - 300 K \cite{Singley2002}. Figure \ref{Fig:singley1} displays the reflectance R($\omega$) below 2,000 \cm measured at different temperatures. The inset shows R($\omega$) up to 50,000 \cm on a logarithmic scale where one can see a plasma edge near 10,000 \cm and an interband transition feature near 17,000 \cm. In nearly the entire spectral range R($\omega$) is above 50$\%$. At room temperature (292 K), R($\omega$) increases monotonically with decreasing frequency.
As the temperature decreases to 100 K, a minimum begins to develop near 500 \cm. This minimum deepens as
the temperature is reduced further, which is an indication for the formation of hybridization energy gap.

\begin{figure}
  \centering
  \includegraphics[width=7cm]{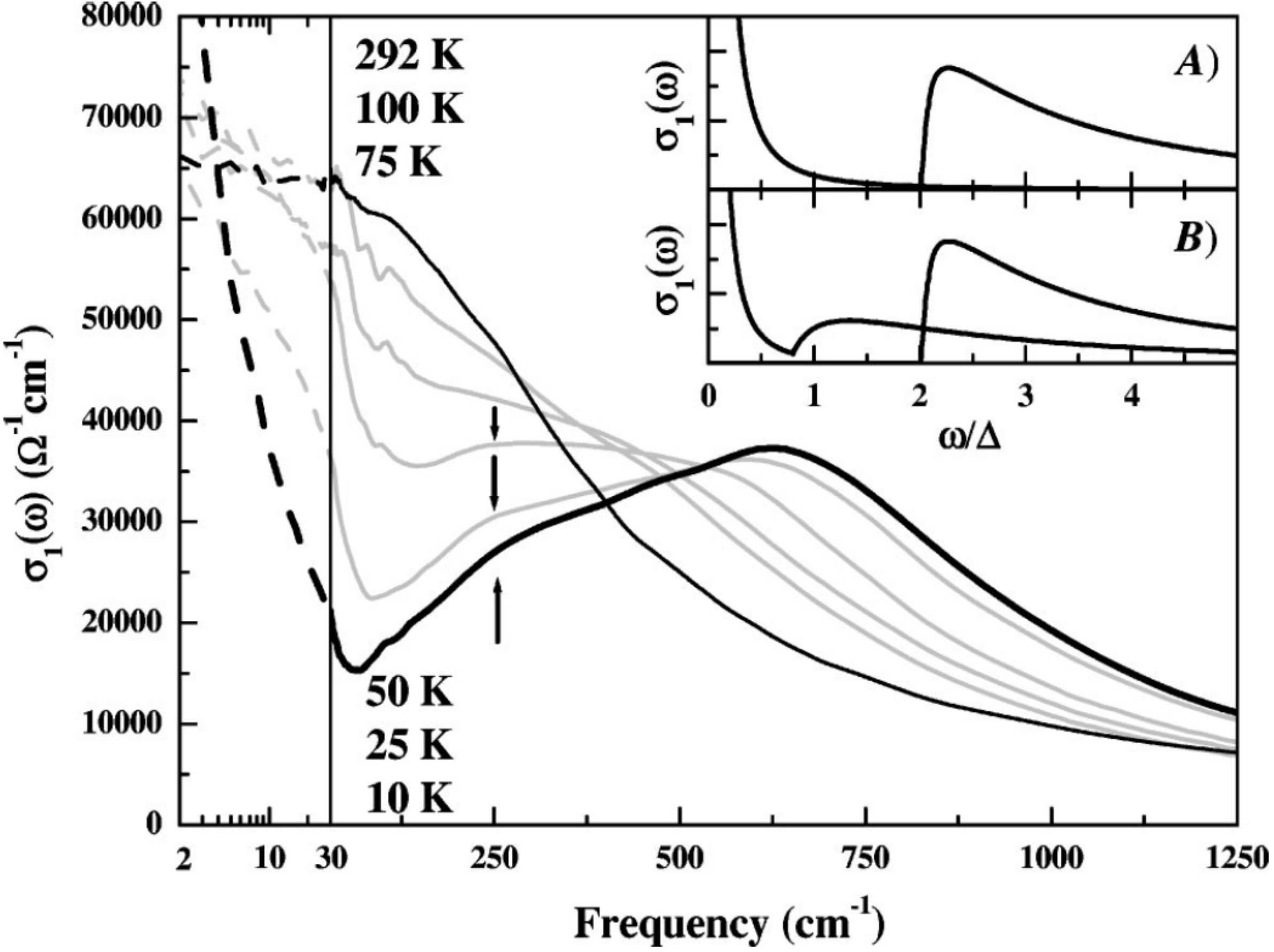}\\
  \caption{Real part of the conductivity $\sigma_1(\omega)$ of CeCoIn$_5$ in the far infrared. At 292 K, $\sigma_1(\omega)$ shows a metallic Drude-like response. At low temperatures the development of a gap in the DOS is apparent from
the peak in $\sigma_1(\omega)$ which grows near 600 \cm. The very narrow
mode at low energies is a signature of the intraband response of the
heavy quasiparticles. The arrows near 250 \cm highlight additional absorption within the gap. Panel A of the insets shows model
calculations of $\sigma_1(\omega)$ for a hybridization gap and heavy quasiparticle
intraband response. In panel B the heavy quasiparticle component is allowed to couple to a boson mode which produces an additional absorption band.\cite{Singley2002}}\label{Fig:singley2}
\end{figure}

The real part of optical conductivity of CeCoIn$_5$ is shown in Fig. \ref{Fig:singley2}. At high temperature,  $\sigma_1(\omega)$ shows normal metallic behavior, with a line shape in accord with Drude theory. As
the temperature is lowered to 100 K, deviations in the spectrum of $\sigma_1(\omega)$ from Drude behavior become evident. In the coherent state at lower temperature, $\sigma_1(\omega)$ exhibits a clear energy gap structure: the spectral weight below 300 \cm is strongly suppressed. It also shows a strong mid-infrared peak centered at $\sim$600 \cm. Meanwhile, a narrow Drude peak develops at very low frequencies, roughly in the extrapolated region. The narrow peak must be present as $\sigma_1(\omega)$, in the low frequency limit, gives rise to the dc transport value.

The above observations in CeCoIn$_5$, including the depletion of spectral weight (i.e. the formation of hybridization energy gap), the development of a mid-infrared peak and the extremely narrow Drude peak, are characteristic and generic features of heavy fermion metals at low temperature\cite{Degiorgi1999,Degiorgi2001}. What is uncommon in CeCoIn$_5$ is the presence of an additional absorption below the hybridization energy gap. From Fig. \ref{Fig:singley2}, one can find a  shoulder near 250 \cm (marked with arrow) in spectra at low temperatures. The authors ascribed the shoulder to the coupling effect of charge carriers to a collective boson mode at the energy of $\sim$65 \cm (8 meV) and suggested an antiferromagnetic origin of the mode. However, similar absorption features are also present in other heavy fermion metals or mixed valent compounds. Such structures could not be easily accounted for on the basis of simple PAM. A discussion on this issue will be presented in the next section.

By optics one can estimate the the mass enhancement of the quasiparticles in the coherent state. The mass enhancement can be estimated by applying a sum rule to the $\sigma_1(\omega)$ data. The spectral weight under the conductivity curve is proportional to the density of carriers divided by their effective mass. Therefore assuming the density
of carriers remains constant and using the fact that m$^*$(292 K)=m$_b$ , one obtains:
\begin{equation}
{{m^*}\over{m_b}}={{\int_0^{\omega_c^{292K}}\sigma_1(\omega,292K)d\omega}\over{\int_0^{\omega_c^{10K}}\sigma_1(\omega,10K)d\omega}}\label{EQ1}
\end{equation}
where $\omega_c^{10K}$ and $\omega_c^{292K}$ are the integration limits at respective temperatures. These integration limits are chosen such that only the intraband component of $\sigma_1(\omega)$ is included
at each temperature (65 \cm at 10 K and 2500 \cm at 292 K).\cite{Singley2002} Following this procedure a value of m$^*$=26m$_b$ is obtained
for the effective mass at 10 K. This enhancement value is within the range of the estimated values based on the thermodynamic\cite{Movshovich2001,Kim2001,Ikeda2001} and de Haas-van Alphen (dHvA) measurements\cite{Settai2001,Shishido2002}.

 \begin{figure}
  \centering
  \includegraphics[width=7cm]{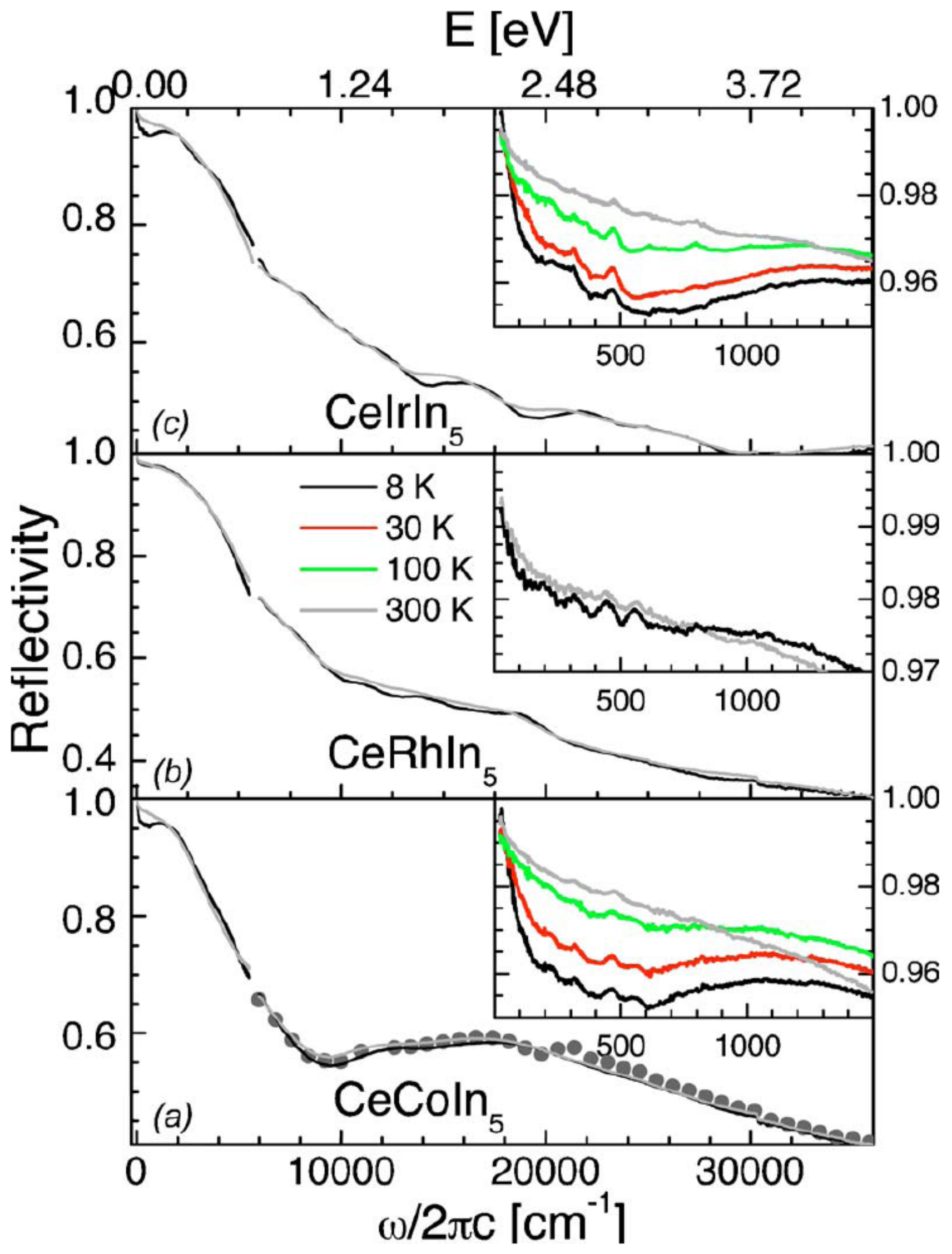}\\
  \caption{Comparison of reflectivity spectra of CeMIn$_5$ (M=Co, Rh, Ir) at several temperatures. The gray dots show the measured reflectivity of the ab plane of CeCoIn5. Insets: Reflectivity below 2000 \cm.\cite{Mena2005}}\label{Fig:mena1}
\end{figure}

 \begin{figure}
  \centering
  \includegraphics[width=7cm]{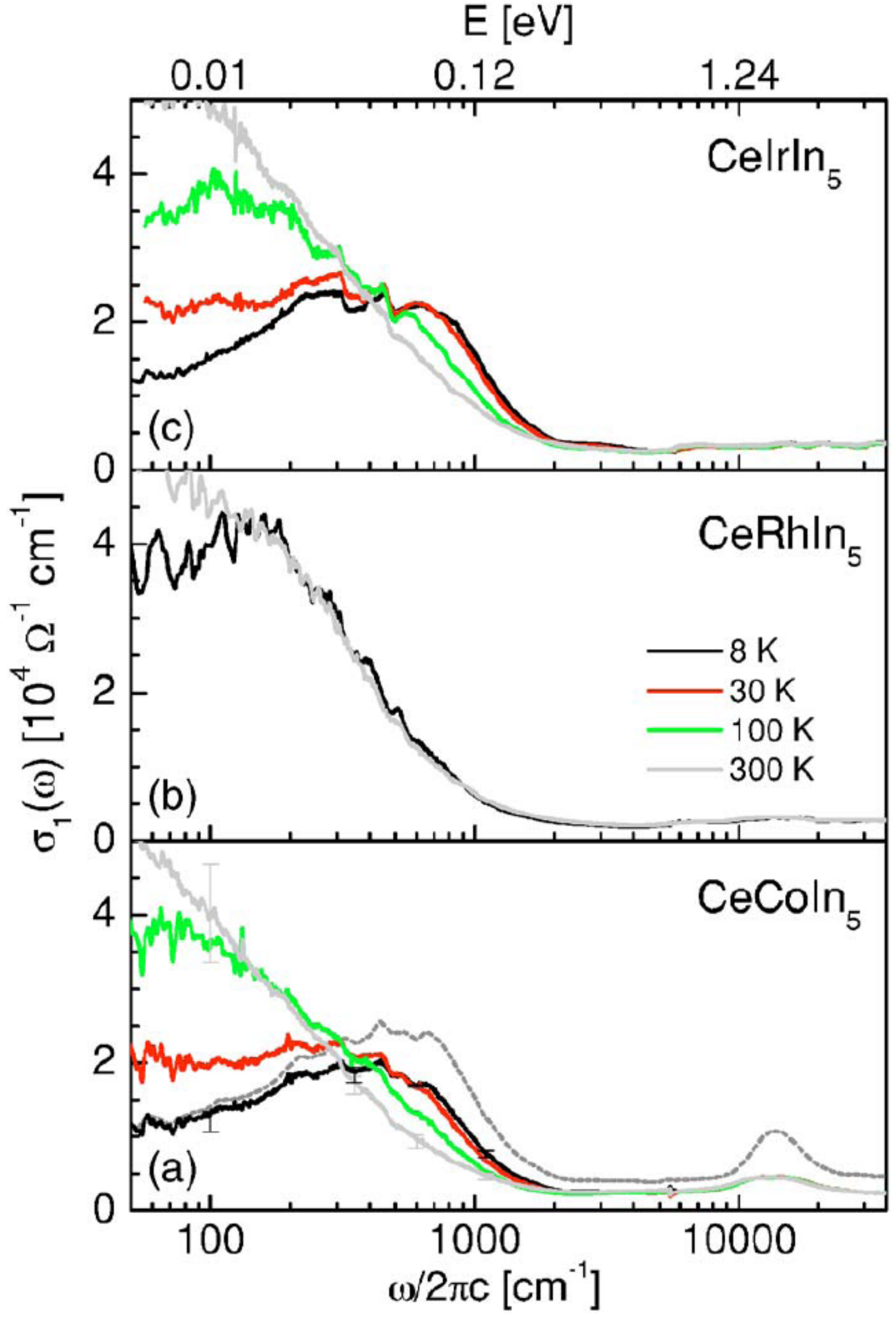}\\
  \caption{Real part of the optical conductivity of CeMIn$_5$ (M=Co, Rh, Ir) at several temperatures. The lower panel also shows the obtained
$\sigma_1(\omega)$ if the high-frequency reflectivity of Ref. \cite{Singley2002} is used in the KK analysis (gray dotted line).\cite{Mena2005}}\label{Fig:mena2}
\end{figure}

Now let us look at two other Ce-based 115 compounds. CeIrIn$_5$ is also a heavy fermion metal and becomes superconducting at $T_C$ =0.4 K \cite{Petrovic2007a}. The optical response of CeIrIn$_5$ is very similar to that of CeCoIn$_5$. This can be seen clearly from the reflectance and conductivity spectra shown in Fig. \ref{Fig:mena1} and Fig. \ref{Fig:mena2}, respectively.\cite{Mena2005} At high temperature (e.g. 300 K), R($\omega$) decreases monotonically with increasing frequency up to roughly 10,000 \cm and a Drude-like peak appears in $\sigma_1(\omega)$. At low temperature below 100 K, a depletion of spectral weight is seen in both R($\omega$) and $\sigma_1(\omega)$, indicating clearly the formation of hybridization energy gap. On the other hand, the midinfrared peak associated with the formation of hybridization gap is rather broad and contains complex structures, which also resembles the response in CeCoIn$_5$.

However, an extremely weak hybridization energy gap feature is seen in CeRhIn$_5$, a 115 compound with AFM order. In the reflectance spectra, the R($\omega$) curve at 8K almost overlaps with the curve at 300 K, as is seen in the middle panel of Fig. \ref{Fig:mena1}. A gap-like suppression of the conductivity is observed only in the very low frequencies, roughly below 100 - 200 \cm. It becomes evident that the hybridization gap structure becomes weaker and its energy scale shifts significantly to lower frequency \cite{Mena2005}.

\subsection{Momentum dependence of the hybridization gap}

As the hybridization gap is a key parameter in the study of heavy fermions, much attention was paid on the proper identification of this gap. According to PAM\cite{Degiorgi1999,Dordevic2001,Hancock2004}, the transitions across hybridization gaps shall present themselves as a very sharp onset at the gap energy $\Delta_{dir}$, and drops smoothly and slowly at higher frequencies, as shown in Fig. \ref{Fig:PAL}. In real materials, the sharp onset will be dispelled due to the finite life time of quasiparticles and turn into a broad peak, which complicates the accurate estimation of the gap energy.
Nevertheless, the optical conductivity of the CeCoIn$_5$ and CeIrIn$_5$, both of which show remarkably broad mid-infrared peaks along with shoulder like features, can not be well explained by PAM. In fact, the shoulder is so intense in the CeIrIn$_5$ compound that the mid-infrared spectrum resembles a double peak structure.
Such observations have also been revealed to different extent in many other heavy fermion or mixed-valence materials, like YbAl$_3$\cite{Okamura2004}, YbCu$_2$Si$_2$\cite{Okamura2007}, YbRh$_2$Si$_2$ and YbIr$_2$Si$_2$\cite{Sichelschmidt2008}. To resolve this problem, Burch \textit{et al.} proposed that these unexpected spectra of hybridization gaps might be associated with the momentum dependence of the hybridization strength, which may account for the diversity in the "115" family as well\cite{Burch2007}.

\begin{figure}[htbp]
  \centering
  \includegraphics[width=7cm]{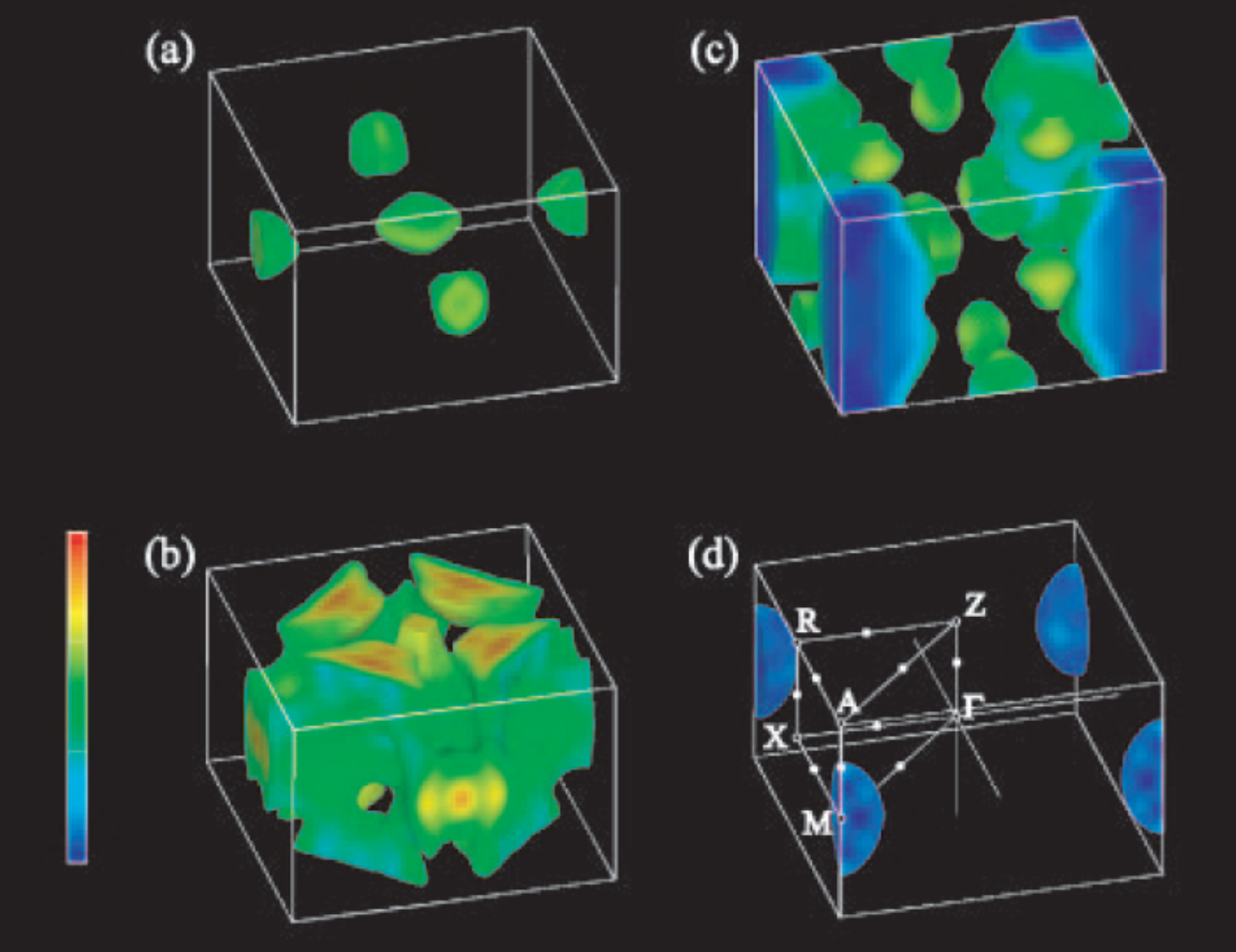}\\
  \caption{Calculated Fermi surface of CeIrIn$_5$. The four sub-image corresponds to the four different bands respectively. The colors indicate the amount of 4$f$ angular momentum character on each sheet of the Fermi surface. Red-shift indicates the increase of the admixture of f electrons. \cite{Maehira2003}}\label{Fig:Maehira1}
\end{figure}

The band structures of CeCoIn$_5$ and CeIrIn$_5$ have been calculated by a relativistic linear augmented-plane-wave method (RLAPW), taking the 4$f$ electrons as itinerant particles.\cite{Maehira2003} The results demonstrated that there are four bands crossing the Fermi energy $E_F$, forming multiple sheets of Fermi surfaces as shown in Fig. \ref{Fig:Maehira1}. Particularly, the Ce 4$f$ and In 5$p$ electrons hybridize with each other in the neighborhood of $E_F$ and participate in the Fermi surface, the degree of which can thus be measured by the contributions of 4$f$ states to the Fermi surface, and is depicted by the color scale in Fig. \ref{Fig:Maehira1}. Obviously, the hybridization changes dramatically from surface to surface, which also varies significantly in the momentum space on the same Fermi surface.

\begin{figure}[htbp]
  \centering
  \includegraphics[width=7cm]{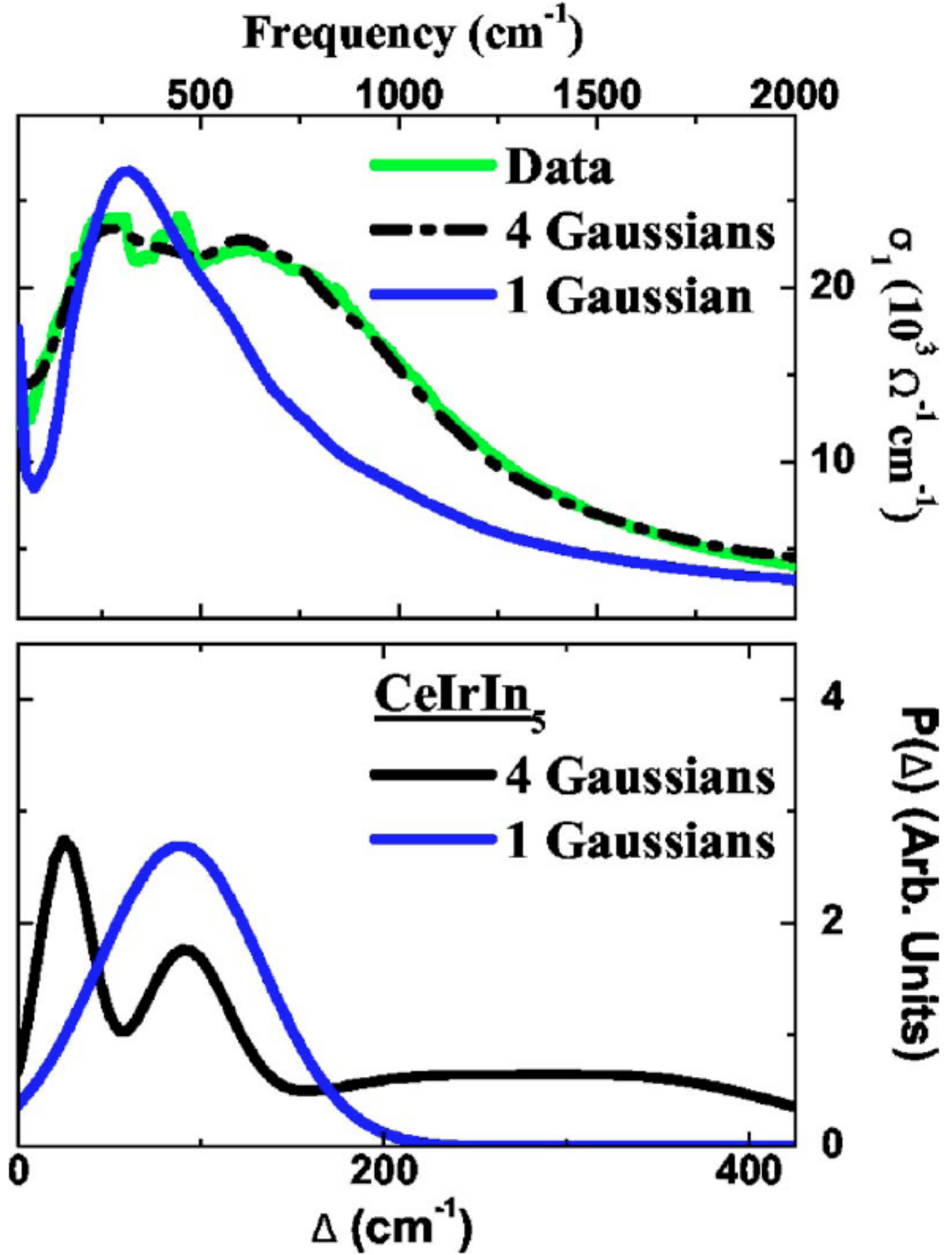}\\
  \caption{The top panel shows experimental optical conductivity of CeIrIn$_5$ in green solid lines. The fitting result when employing one and four Gaussians are plotted by blue and black lines, respectively. The resulting distribution of hybridization gaps $P(\Delta)$ is displayed in the bottom panel.\cite{Burch2007} }\label{Fig:kdependence}
\end{figure}

The band structure calculation provided a solid foundation for Burch's proposal. The strong momentum dependence of hybridization strength will naturally leads to a momentum dependent distribution of the hybridization gap value $\Delta$. Assuming the assignment of $\Delta$ could be described by a function $P(\Delta)$, the optical conductivity $\sigma_1^{PAM}$ confined to the PAM can be modified to a convolution with $P(\Delta)$:
\begin{equation}\label{Eq:Burch1}
  \sigma_1(\omega)=\int_0^{\omega_C}P(\Delta)\sigma_1^{PAM}(\omega,\Delta)d\Delta.
\end{equation}
It is too difficult to resolve the exact form of $P(\Delta)$ by fitting the experimental data with this formula, mainly due to the integration. Nevertheless, $P(\omega)$ could be simply expressed as the sum of several Gaussian functions, modeling the momentum dependence of $\Delta$. The result of only one Gaussian function is plotted in the upper panel of Fig. \ref{Fig:kdependence}, which fails to reproduce the experimental optical conductivity of CeIrIn$_5$. However, it does provide some implications about how the gap value distribution affects the optical conductivity. It is clearly seen in the lower panel of Fig. \ref{Fig:kdependence} that a hybridization gap with energy within 200 \cm can influence the optical spectrum well above this frequency. It is found that four Gaussians are required to generate the primary characters shown in the spectrum of CeIrIn$_5$, in agreement with the band structure calculation.

\begin{figure}[htbp]
  \centering
  \includegraphics[width=7cm]{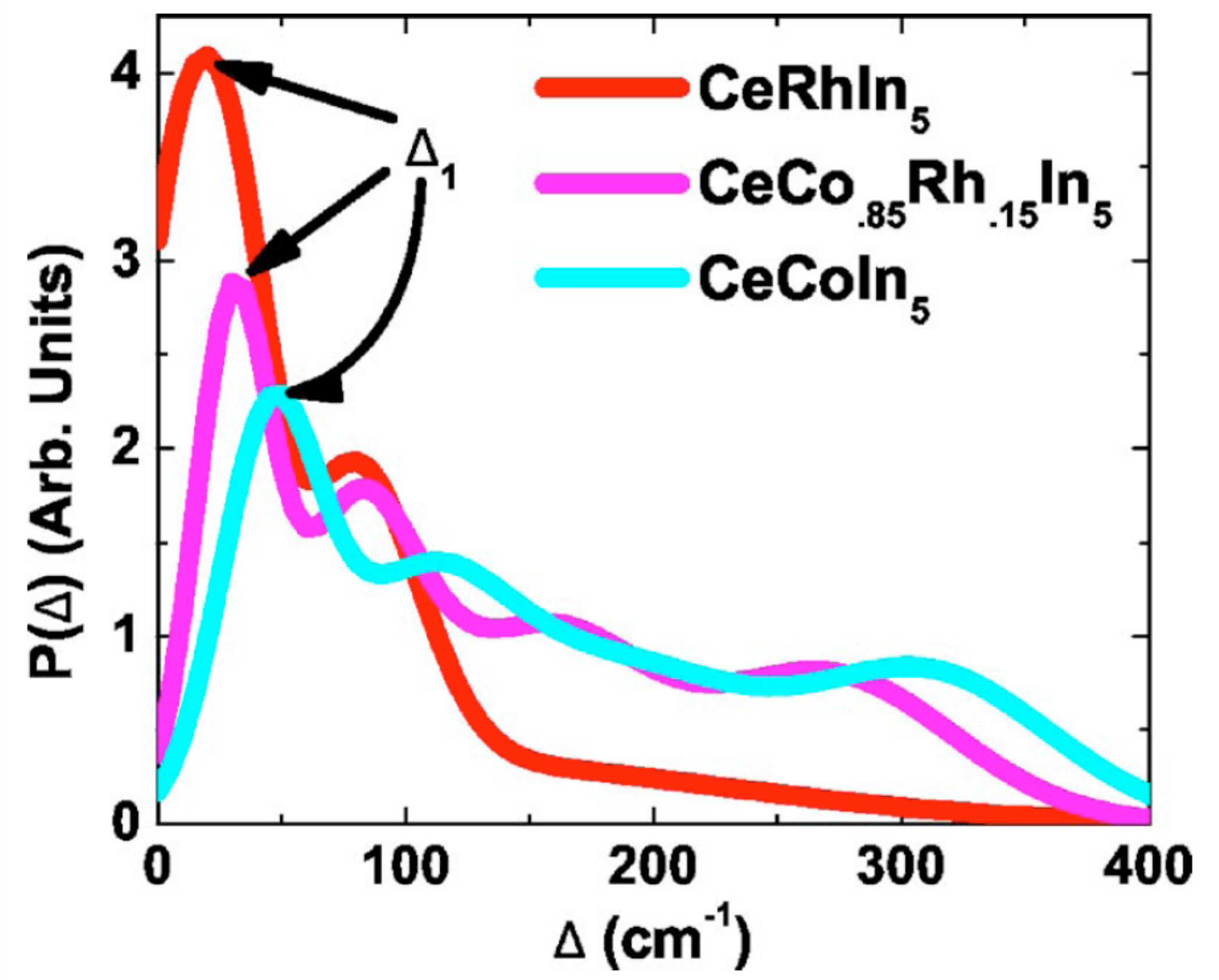}\\
  \caption{The spectrum of hybridization gap values for CeRhIn$_5$, CeCo$_{0.85}$Rh$_{0.15}$In$_5$ and CeCoIn$_5$. \cite{Burch2007} }\label{Fig:burch2}
\end{figure}

The obtained hybridization gap distribution of CeRhIn$_5$, CeCo$_{0.85}$Rh$_{0.15}$In$_5$ and CeCoIn$_5$ are displayed in Fig. \ref{Fig:burch2}. The relatively broad peak centered between 300-400 \cm in $P(\Delta)$ is responsible for the shoulder arising in the optical conductivity of CeCoIn$_5$ and CeIrIn$_5$. A common character of all the four compounds (together with CeIrIn$_5$ in the bottom panel of Fig. \ref{Fig:kdependence}) is that the first Gaussian peak with the lowest energy is of the strongest intensity, the central position of which is labeled as $\Delta_1$, standing for the average gap value on a particular band. When replacing Co with Rh, $\Delta_1$ keeps moving to lower energy, in correspondence with the decreasing of $\tilde{V}$.

Another prominent feature of the gap distribution function is that $P(\Delta)$ does not vanish when $\Delta$ approaches zero. In other words, there are nodal regions in the momentum space, which means the local moments are not totally screened by the conduction electrons. This is in accordance with the intensively discussed two-fluid model in recent year\cite{Nakatsuji2004,Yang2008,Shirer2012a,Shockley2013,Jiang2014}. Especially, the $P(\Delta=0)$ value of CeRhIn$_5$ is much larger than that of CeCoIn$_5$ and CeIrIn$_5$, in agreement with its antiferromagnetic ground states. Moreover, the appearance of node also permits the existence of magnetic fluctuations, which is believed to mediate the superconducting paring mechanism.

The above procedure and analysis clarify that the discordance between the simple PAM prediction and experimental results of CeMIn$_5$ compounds could be well explained by the momentum dependence of multiple bands taking part in hybridizations. Later on, Shim \emph{et al}. calculatee the optical conductivity of CeIrIn$_5$ at different temperatures, utilizing first principle dynamical mean field theory (DMFT) in combination with local density approximation (LDA).\cite{Shim2007} The obtained results coincide with the experimental data qualitatively, including a broad Drude peak at room temperature, the narrowing of Drude peak at extremely low temperatures and the double peak structure in the mid-infrared region. The first-principles LDA+DMFT treatment places the momentum-dependent hybridizations in a microscopic framework.

\begin{figure}[htbp]
  \centering
  \includegraphics[width=8cm]{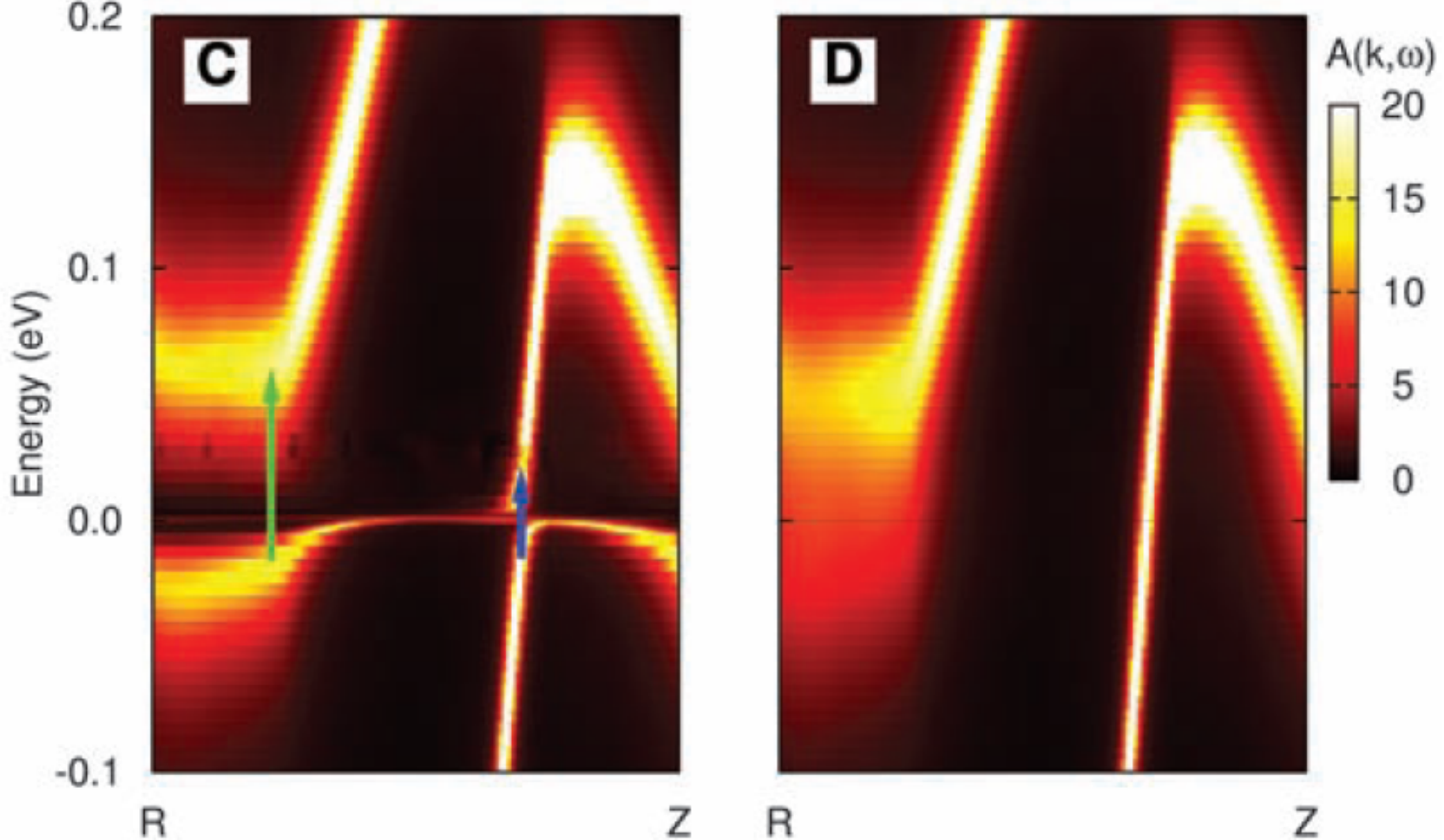}\\
  \caption{The momentum-resolved conduction electron spectral function at 10 K (C) and 300 K (D). \cite{Shim2007} }\label{Fig:shim1}
\end{figure}

With this method, the spectral function $A(\textbf{k}, \omega)$ which describes the quantum mechanical probability of removing or adding an electron with momentum $\textbf{k}$, and energy $\omega$ could also be yielded. The left and right panel of Fig. \ref{Fig:shim1} display the non-Ce 4$f$ spectral function ($A_{total}-A_{Ce}$) along a high symmetry line at 10 K and 300 K respectively. Namely, only the conduction electrons are involved in this print. It is clearly seen that there are huge discrepancies when temperature changes. At 300 K, two bands with very strong dispersions are exhibited, which mainly stem from the In 5p orbitals. Specifically, the left band corresponds to In atoms in the IrIn$_2$ layer (out of plane), whereas the right band origins from In atoms in the CeIn$_3$ layer (in plane). When temperature is lowered to 10 K, both of the two bands hybridize with the very flat f band and two gaps of different energies opened up near $E_F$. This result gives strong backing to the multiple gaps scenario.

Surprisingly, the hybridization gap due to the out of plane In atom is much larger than that of the in plane In atom, indicating that Ce 4$f$ band preferably hybridizes with the out of plane In atom rather than the in plane one. It is very likely that the transition metals in the CeMIn$_5$ compounds couple with In orbitals as well, which further influence the Ce-In hybridization strength. This speculation highlights the role of transition metals and hints why the substitution of transition metals gives rise to different ground states.

\subsection{Infrared properties of CeIn$_3$ and CePt$_2$In$_7$}
As discussed in the Introduction, when the coupling strength $J$ is very small, the RKKY interaction triumphs and the local moments tend to form a long range antiferromagnetic order with decreasing thermal fluctuation, corresponding to the heavy fermion antiferromagnets. Besides several members in the Ce$_m$M$_n$In$_{3m+2n}$, examples are also found in other compounds, such as CeCu$_2$Ge$_2$\cite{Jaccard1992}, CeRh$_2$Si$_2$\cite{Movshovich1996}, and Ce$_2$Ni$_3$Ge$_5$\cite{Hossain2000} ect. In this case, the magnetic susceptibility experiences a paramagnetic to antiferromagnetic phase transition at the Neel temperature $T_N$. Although the ground state of this type materials seems not to be intimately relevant to heavy fermion state, their specific heat usually shows a significant enhancement at low temperature, even below $T_N$, which is identified to stem from a large effective mass instead of fluctuations. The magnetic susceptibility deviates from the Curie-Weiss law at temperatures well above $T_N$, and the resistivity regularly shows downward curvature in the meanwhile, both of which are strong implications of Kondo screening.

\begin{figure}[htbp]
  \centering
  \includegraphics[width=7cm]{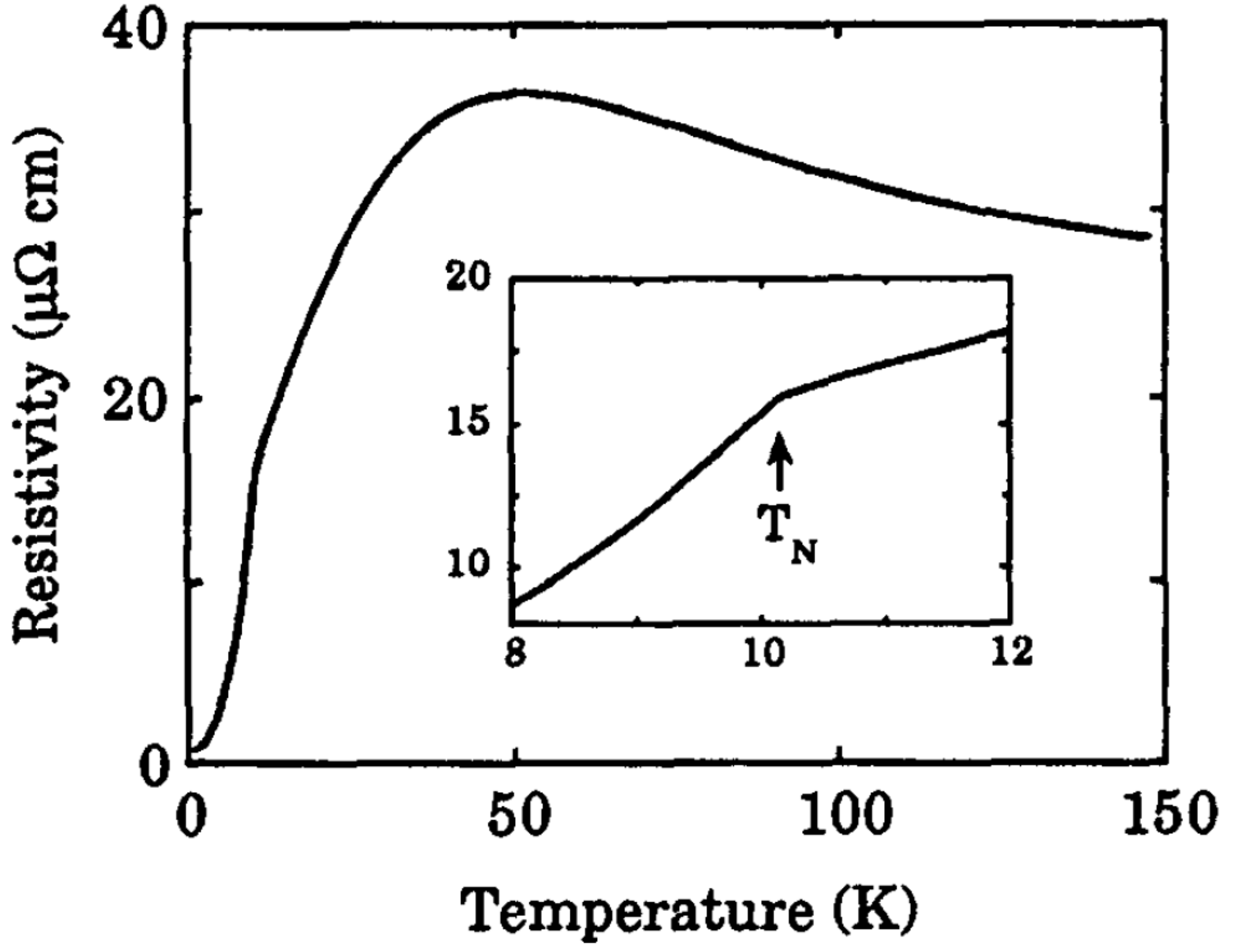}\\
  \caption{The temperature dependent resistivity of CeIn$_3$. The inset is an expanded view in the vicinity of the antiferromagnetic transition temperature near 10 K.\cite{Walker1997}}\label{Fig:walker1}
\end{figure}

As a parent compound of the Ce$_m$M$_n$In$_{3m+2n}$ family, the CeIn$_3$ single crystal is of AuCu$_3$-type cubic structure, which experiences an antiferromagnetic phase transition at $T_N$ = 10.1 K.\cite{Walker1997} As can be seen in Fig. \ref{Fig:walker1}, the resistivity of CeIn$_3$ shows a broad peak around 50 K and a cusp feature near 10 K, the latter of which is a hallmark of the formation of antiferromagnetic order. The maximum in the temperature dependent resistivity is a common character in heavy fermion systems, which directly bounds up with the Kondo coherence, signaling the onset of local momentum screening.

\begin{figure}[htbp]
  \centering
  \includegraphics[width=7cm]{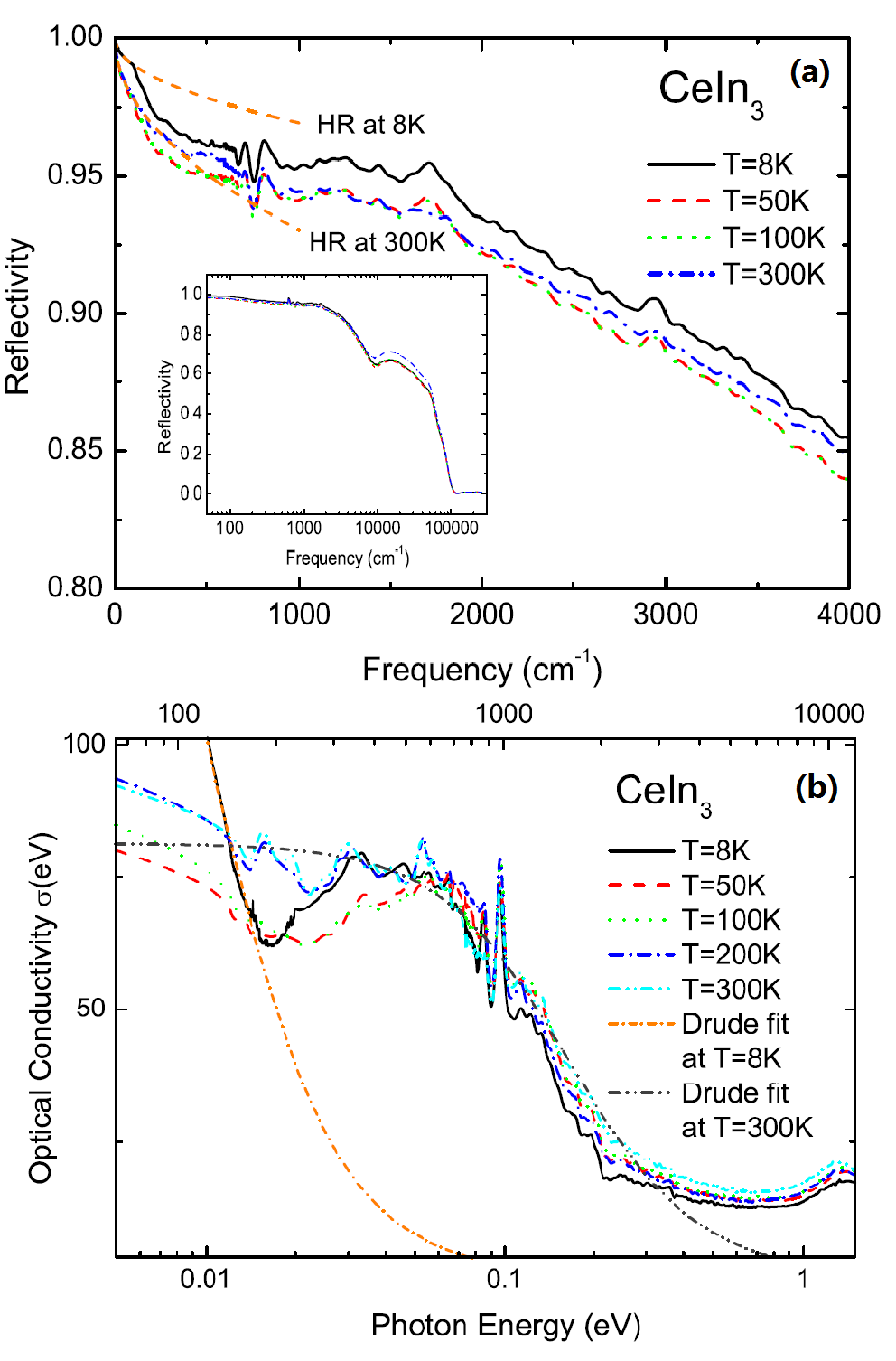}\\
  \caption{Panel A: the frequency dependent reflectivity of CeIn$_3$ at several selected temperatures. The orange dotted lines are reflectivity calculated from a Hagen-Rubens' relation. The inset shows $R(\omega)$ in a whole measured range. Panel B: the real part of optical conductivity of CeIn$_3$ at selected temperatures.\cite{Lee2008}}\label{Fig:lee1}
\end{figure}

The frequency dependent reflectivity of this compound was initially measured by Lee \emph{et al}.\cite{Lee2008}, as shown in the upper panel of Fig. \ref{Fig:lee1}. For all the selected temperatures, $R(\omega)$ exhibits metallic behavior. At 8 K, $R(\omega)$ drops rapidly at very low energy then becomes quite flat above 500 \cm, while it decreases quickly again beyond 1800 \cm. The shape of reflectivity is much alike depletion due to interband transitions, resembling the reflectivity of CeMIn$_5$. However, the spectrum at 300 K shows a very similar manner with 8 K, which is quite different from that of CeMIn$_5$ at room temperature. The orange dotted lines in the main panel of Fig. \ref{Fig:lee1} are Hagen-Rubens predictions of $R(\omega)$, which coincides well with the experimental data at 300 K, but deviates from the data at 8 K. This behavior hinted the development of hybridization energy gap.

The real part of optical conductivity is displayed in the lower panel of Fig. \ref{Fig:lee1}. At room temperature, the low energy response could be roughly reproduced by the metallic Drude fitting with a broad half width, except for a dip feature around 250 \cm. Since this dip exists all over the measured temperature range, it is most likely to be some kind of artifact. As a contrast, the Drude peak of 8 K substantially shifts to lower frequencies, and the half width of it gets much smaller. This observation is consistent with the Kondo coherence of heavy fermions. In addition, a broad peak emerges at low temperatures in the mid-infrared range, which is supposed to be attributed to excitations across the hybridization gap. The magnetic ground state of CeIn$_3$ implies that the RKKY strength wins out in the competition with Kondo interaction. In this respect, a weak and low energy hybridization gap is expected. However, it is difficult to identify the gap energy due to the poor data quality.

Iizuka \emph{et al.} also performed optical measurements on CeIn$_3$ single crystals\cite{Iizuka2012} and their optical conductivity seems to be more reasonable. Above 30 K, $\sigma_1(\omega)$ increases monotonically with decreasing energy and can be well accounted for by the metallic Drude model, without any extra absorption. Below 30 K, a very weak mid-infrared peak gradually develops at about 20 meV. In the meanwhile, the spectral weight of the Drude peak is dramatically reduced, due to the formation of resonance of heavy quasiparticles. This observation agrees well with the drop of resistivity below 50 K. Comparing with the 115 family, the hybridization gap identified here resembles the spectra of CeRhIn$_5$, which is much weaker than that of CeCoIn$_5$ and CeIrIn$_5$, in accordance with the Doniach phase diagram.

\begin{figure}[htbp]
  \centering
  \includegraphics[width=7cm]{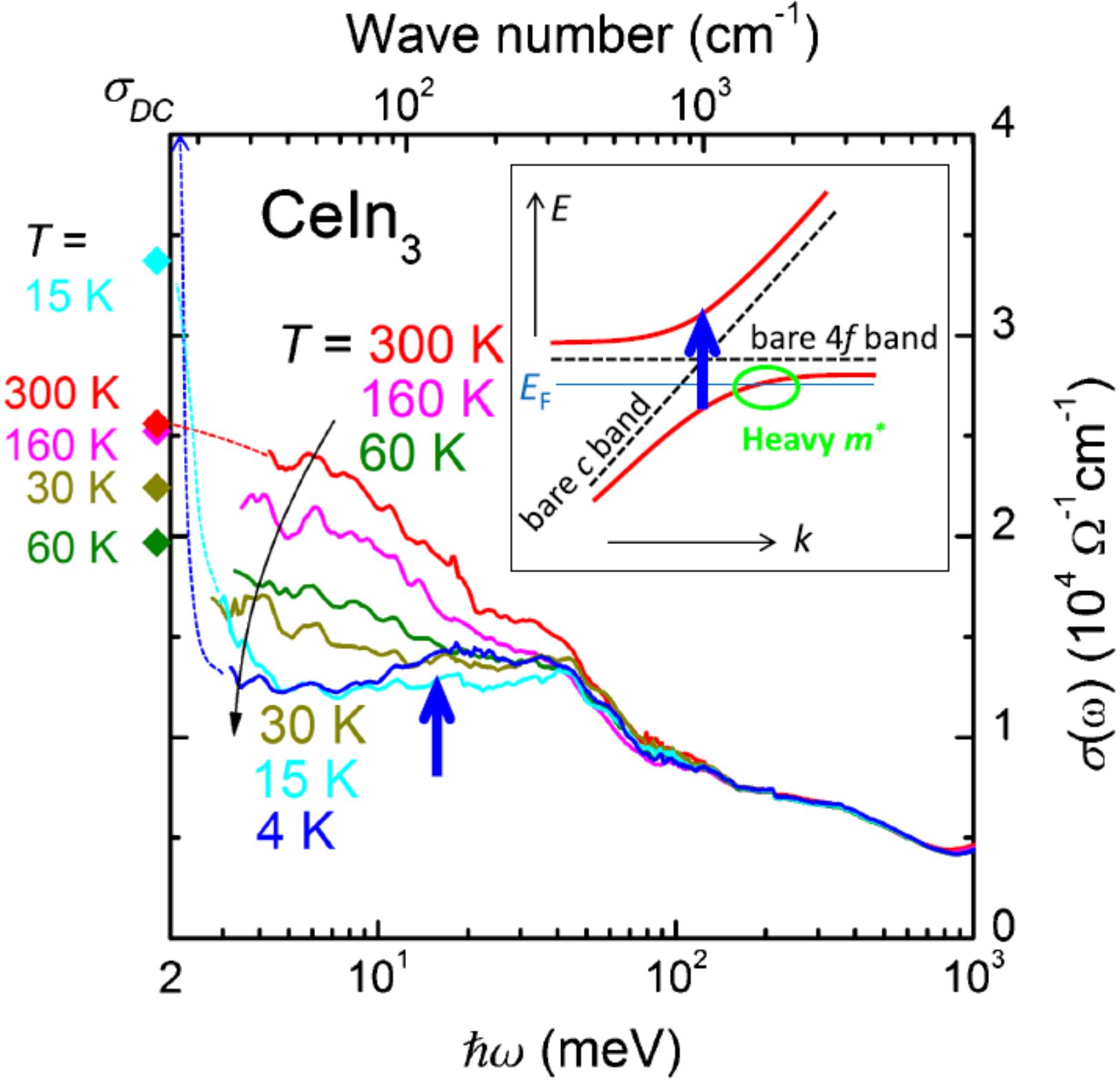}\\
  \caption{The optical conductivity $\sigma_1(\omega)$ of CeIn$_3$. The solid diamonds show the direct current resistivity $\sigma_{DC}$ at different temperatures. The dotted lines are extrapolations of $\sigma_1(\omega)$ spectra to $\sigma_{DC}$. The inset shows a sketch of \emph{c-f} hybridization. \cite{Iizuka2012}}\label{Fig:iizuka1}
\end{figure}

Apart from that, there has been a longstanding debate on whether the c-f hybridization exists (i.e. Kondo breakdown) when the long range magnetic order sets in\cite{Paschen2004,Shishido2005,Settai2005,Kadowaki2006,Watanabe2008a,Friedemann2010a,Watanabe2010}. CeIn$_3$ is an ideal material to address this issue, owing to its relatively high Neel temperature. Here, the optical conductivity was measured as low as 4 K (well below $T_N$=10.2 K), which illustrated the existence of itinerant heavy fermions deep in the antiferromagnetic state. In addition, with the application of pressure, both the energy and intensity of the hybridization peak increase continuously, as the ground state was tuned from AFM phase to the paramagnetic regime\cite{Iizuka2012}.

Next, we will discuss the optical properties of a more two dimensional compound CePt$_2$In$_7$, as compared with the CeIn$_3$ and CeMIn$_5$ systems. Generally speaking, reduced dimensionality is believed to benefit unconventional superconductivity, since it increases the probability of nesting type instabilities. For example, the iron-based and cuprate superconductors are all two dimensional materials and the unconventional superconductivity only occures in the FeAs and CuO layers. Consequently, people tried to seek for higher superconducting transition temperatures in CePt$_2$In$_7$. Nevertheless, the ground state of CePt$_2$In$_7$ compound was determined to be antiferromagnetic instead of superconductivity, with a Neel temperature $T_N$ = 5.2 K\cite{ApRoberts-Warren2010}, although there is no consensus on whether the long range antiferromagnetic order is commensurate or incommensurate so far\cite{ApRoberts-Warren2010,Sakai2014,Mansson2014a}.

\begin{figure}[htbp]
  \centering
  \includegraphics[width=7cm]{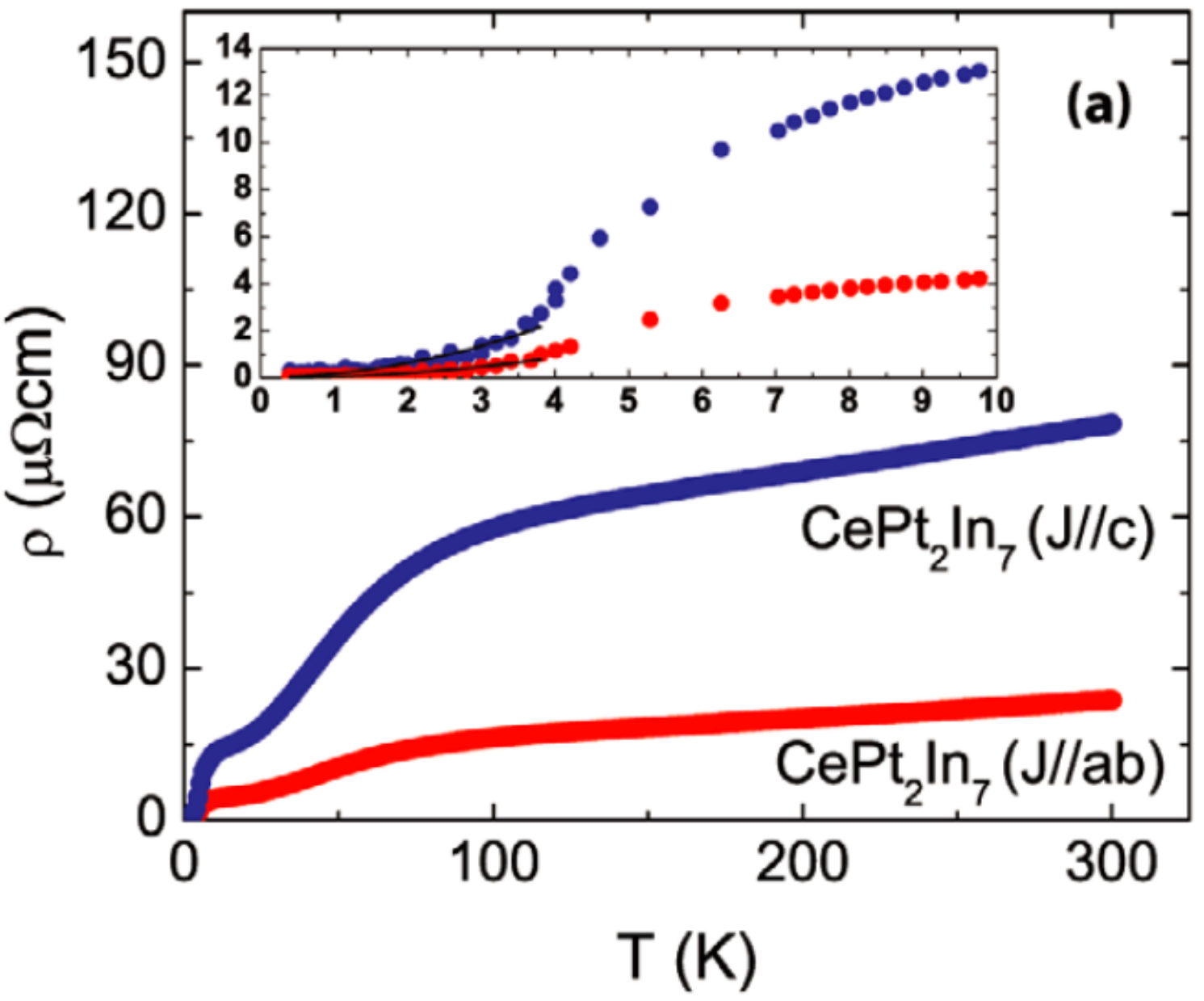}\\
  \caption{The resistivity of CePt$_2$In$_7$ as a function of temperature along two different orientations. The inset shows an expanded view from 0.4 K to 10 K. \cite{Tobash2012}}\label{Fig:tobash1}
\end{figure}

The temperature dependent resistivity of CePt$_2$In$_7$ is displayed in Fig. \ref{Fig:tobash1}, which keeps decreasing as temperature decreases and shows a gentle downturn below 80 K. The overall profile is similar to the resistivity of CeIn$_3$ as shown in Fig.\ref{Fig:walker1}, and the inflection around 80 K is therefore ascribed to the screening of local magnetic moments.
Moreover, the specific heat measurement reveals a very large value of the Sommerfeld coefficient above the Neel temperature\cite{Bauer2010}, verifying the existence of heavy fermions associated with Kondo effect. It seems that the CePt$_2$In$_7$ compound locates at the same position as
CeIn$_3$ and CeRhIn$_5$ in the Doniach phase diagram, in spite of the more two dimensional Fermi surface\cite{Bauer2010}.

\begin{figure}[htbp]
  \centering
  \includegraphics[width=7cm]{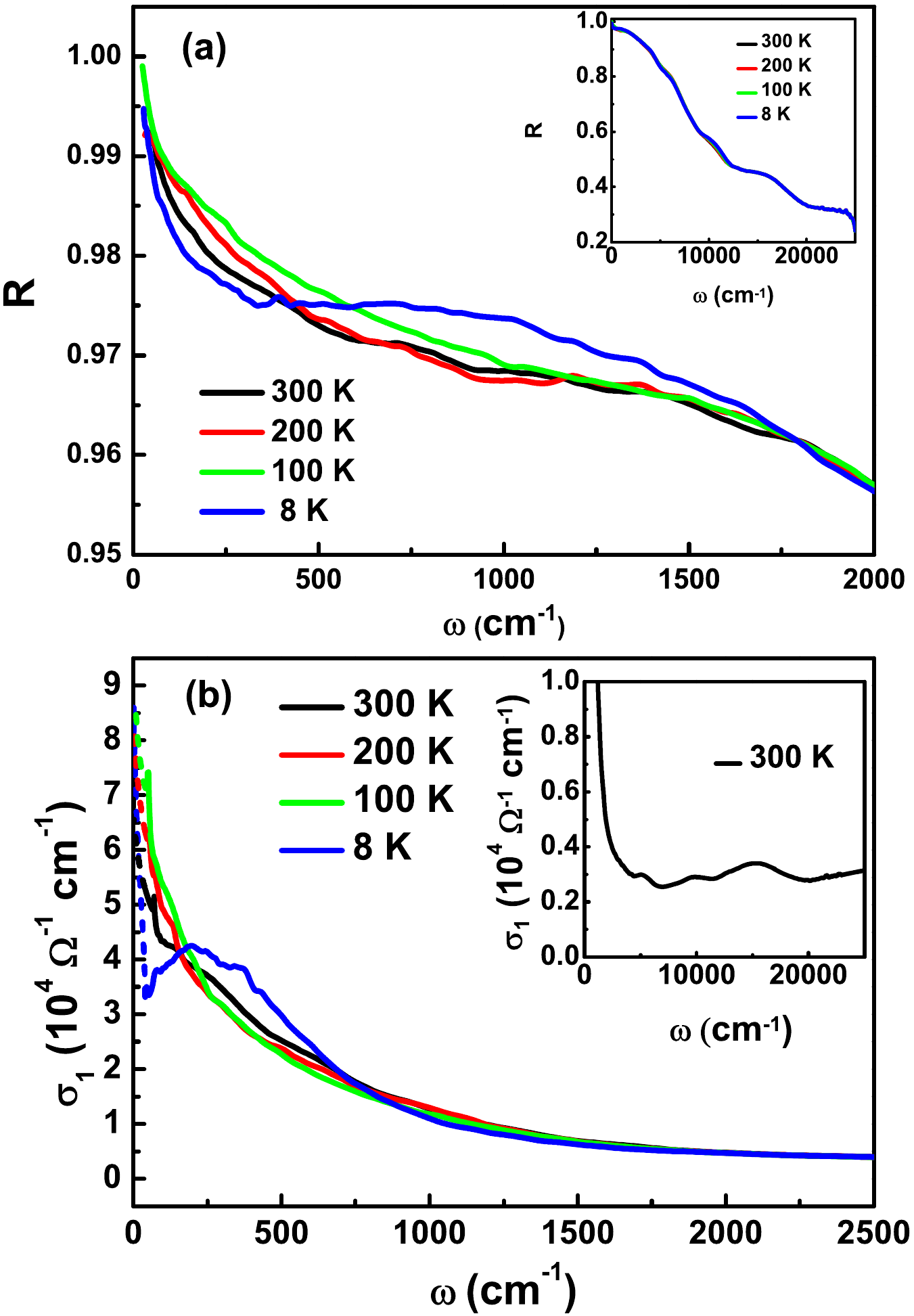}\\
  \caption{The frequency dependent (a) reflectivity and (b) optical conductivity of CePt$_2$In$_7$ at several selected temperatures.\cite{Chen}}\label{Fig:R127}
\end{figure}

The frequency dependent reflectivity $R(\omega)$ and optical conductivity ($\sigma_1(\omega)$) of the single crystalline CePt$_2$In$_7$ are plotted in the upper and lower panel of Fig. \ref{Fig:R127}, separately\cite{Chen}. In analogy with CeIn$_3$ and CeRhIn$_5$, the reflectivity shows metallic behavior all over the measured temperature range. In the far infrared region below 500 \cm, the value of $R(\omega)$ increases monotonically with temperature cooling, coinciding with the monotonic decrease of resistivity. Exceptionally, reflectivity of the lowest temperature 8 K is slightly suppressed and gets lower than the high temperature reflectivity. In correspondence, the optical conductivity above 100 K increases continuously as the frequency decreases, a typical Drude-like behavior, whereas $\sigma_1(\omega)$ at 8 K exhibits a resonance peak around 190 \cm, as a result of excitations across the hybridization gap. Notably, the Drude peak survives even at 8 K, because the extrapolation of $\sigma_1(\omega)$ to zero frequency should be very large, being consistent with the extremely low dc resistivity.
It is just too sharp to show in the current energy scale.

To summarize, both of CeIn$_3$ and CePt$_2$In$_7$ compounds sit in the antiferromagnetic region of the Doniach phase diagram. Their optical conductivities exhibit very weak hybridization peaks and the energy scales of such peaks are much smaller than the paramagnetic heavy fermion metals like CeCoIn$_5$ and CeIrIn$_5$. In the meantime, the Drude peaks shift drastically to lower frequencies in associate with the formation of heavy quasiparticles. These phenomena established some general trends of the electrodynamic response of a class of heavy fermion antiferromagnets.

\subsection{The parameters that contribute to hybridization strength}
\begin{figure}[htbp]
  \centering
  \includegraphics[width=8cm]{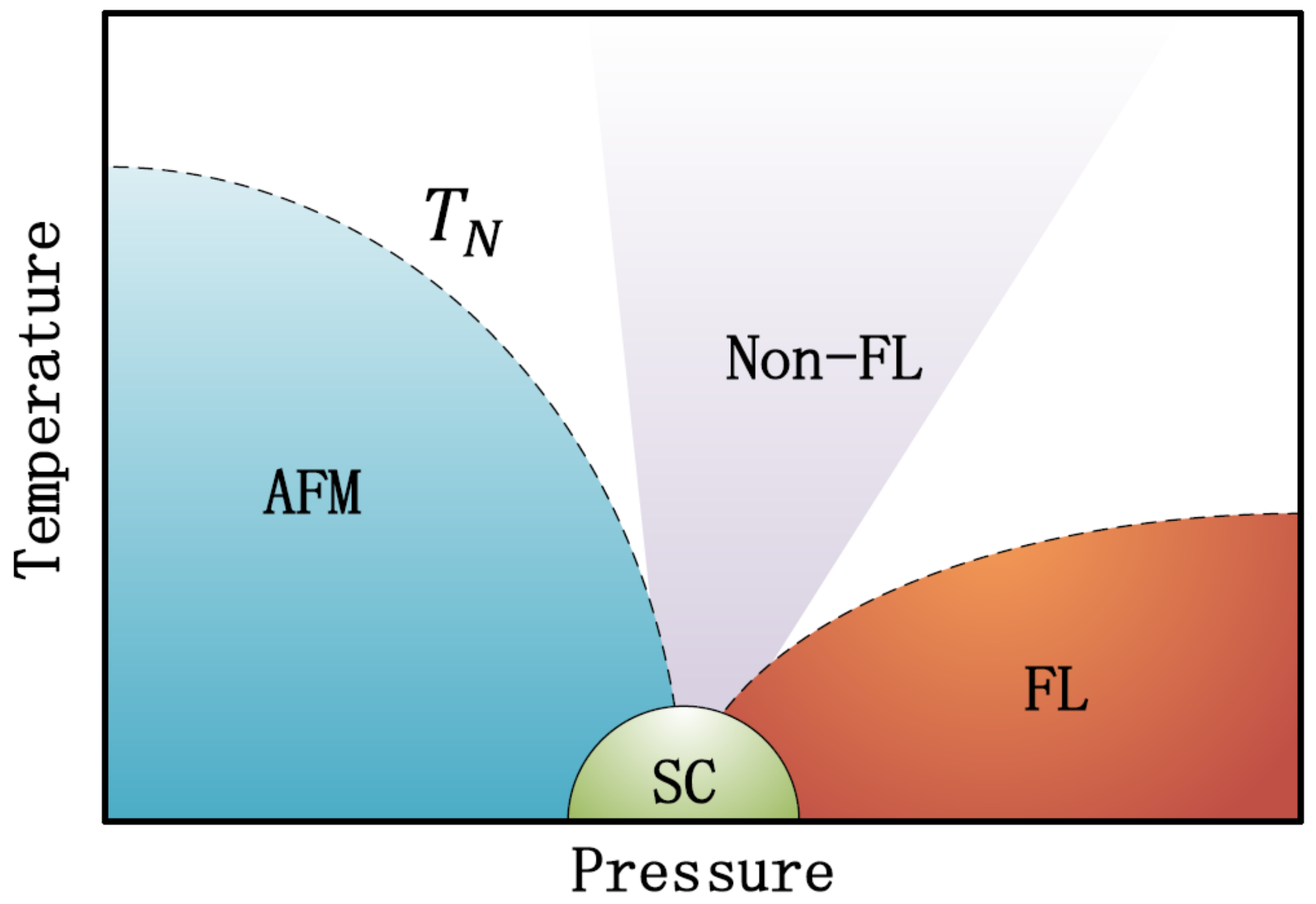}\\
  \caption{A temperature versus pressure phase diagram for heavy fermion systems. The ground state properties could be continuously tuned by external pressure in a way similar to the Doniach phase diagram.}\label{Fig:pressure}
\end{figure}

The coupling strength $J$ is proportional to $\tilde{V}^2/U$, where $\tilde{V}$ is the strength of hybridization, and $U$ represents the Coulomb repulsion in the $f$ band\cite{Schrieffer1966,Cornut1972,Vekic1995}. Apparently, the Doniach phase diagram is continuously tunable by pressure, doping or magnetic field. The application of pressure normally leads to the shrinkage of the lattice constant, which thereby increases the overlap between wave functions of the hybridizing $f$ and conduction electrons, beneficial to a larger $\tilde{V}$\cite{Wills1983}. For heavy fermion antiferromagnets, the long range magnetic order could be gradually suppressed by increasing pressure and gives way to superconductivity in the vicinity of quantum critical point. A simple universal PT phase diagram is plotted in Figure \ref{Fig:pressure}, which crudely reproduced the Doniach phase diagram, although some details concerning the coexistence of different orders are neglected. For heavy fermion paramagnetic metals like CeCoIn$_5$, which sit near the QCP with an intermediate coupling strength and show non-Fermi liquid behaviors, the application of pressure will suppress the magnetic fluctuations and drive the compounds into Fermi liquid state. Although the Doniach picture gives a good description on heavy fermions, the real materials are far more complicated and all the delicate parameters like lattice constant, dimensionality, crystal symmetry and crystal field effect may play an essential role in determining the coupling strength. Consequently, even isostructural compounds with similar lattice constants could exhibit totally different ground states.

It is clearly seen that the hybridization strength $\tilde{V}$ alters dramatically in the Ce$_m$M$_n$In$_{3m+2n}$ family. The similarities and systematic evolutions borne in these compounds provided a ideal platform for the exploration on the contributions of different parameters to $\tilde{V}$. In the following, we will discuss all the potential factors which may contribute to the hybridization strength.

\subsubsection{Lattice structure}

Lattice structure is one of the most fundamental properties of a compound under research, which is directly related to the electronic structure. Especially, the variation of lattice constant caused by elements substitution changes the chemical pressure effectively.
In the Ce$_m$M$_n$In$_{3m+2n}$ family, all the materials possess either tetragonal (CeMIn$_5$ and CePt$_2$In$_7$) or cubic (CeIn$_3$) structures. Particularly, the three members of CeMIn$_5$ hold nealy identical structures. Therefore it was expected that the hybridization strength $\tilde{V}$ would continuously increase as the transition metal iron in the CeMIn$_5$ system changes from Ir to Rh then Co, since the radii of them keep decreasing in the process, which was supposed to enhance the chemical pressure. Nevertheless, the energy scale of hybridization gaps of CeCoIn$_5$ and CeIrIn$_5$ identified by optical conductivity are much larger than that of CeRhIn$_5$, which is quite confusing.

We have summerized the lattice parameters of these compounds in Table \ref{1}. It is noted that the lattice constant of the $a$ axis is very close to each other, indicating that the CeIn$_3$ block they have in common stays in similar environments. The piling of CeIn$_3$ and MIn$_2$ blocks produced a huge difference in the $c$ axis lattice constant. In Table \ref{1}, the value of $a$ decreases monotonically from top to bottom, in accordance with the predicted order of CeMIn$_5$ compounds through the radii of transition metals, whereas $c$ decreased in the opposite trend. The reverse variation of lattice constants makes it difficult to predict the evolution of $\Tilde{V}$ in these materials merely by the respect of overall chemical pressure.

\begin{table}[htbp]
\setlength\abovecaptionskip{0.5pt}
\caption{The lattice parameters and phase transition temperatures at ambient pressure of the Ce$_m$M$_n$In$_{3m+2n}$ family\cite{Moshopoulou2002a,Mansson2014a}.  \label{1}}
\vspace{-1em}
\begin{center}
\renewcommand\arraystretch{1.5}
\begin{tabular}{p{1.5cm} p{1.2cm} p{1.2cm} p{1.2cm} p{1.2cm} p{1cm}}
\hline
\hline
 &$a({\AA})$&$c({\AA})$&$c/a$&$T_c(K)$&$T_N(K)$\\
\hline
CeIn$_3$&4.689&4.689&1&&10\\
CeIrIn$_5$&4.674&7.501&1.605&0.4&\\
CeRhIn$_5$&4.656&7.542&1.620&&3.8\\
CeCoIn$_5$&4.613&7.551&1.637&2.3&\\
CePt$_2$In$_7$&4.609&21.627&4.692&&5.2\\
\hline
\hline
\end{tabular}\\
\end{center}
\end{table}

\begin{figure}[htbp]
  \centering
  \includegraphics[width=8.5cm]{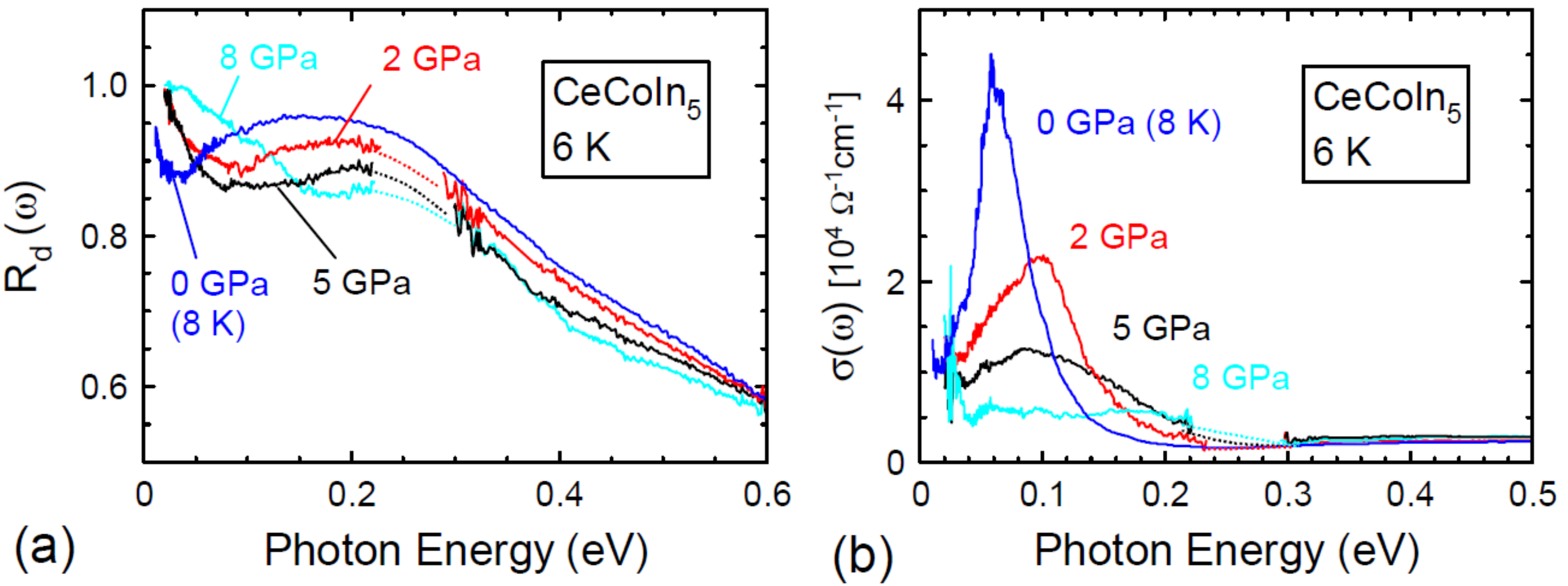}\\
  \caption{Optical reflectance [$R_d(\omega)$] (a) and conductivity [$\sigma(\omega)$] (b) of CeCoIn$_5$ measured at high pressure and at low temperatures. The broken curves bridged by dotted lines are the spectral range could not be measured due to strong absorption by the diamond anvil.\cite{Okamura2015}}\label{Fig:okamura3}
\end{figure}

Okamura \emph{et al}. have investigated the optical responses of CeCoIn$_5$ with application of pressure along the axial $c$ axis\cite{Okamura2015}. The evolution of reflectivity $R_d(\omega)$ under different pressures is displayed in the left panel of Fig. \ref{Fig:okamura3}. The depletion feature below 0.1 eV under ambient pressure continually shifts to higher energies and gets much broader as the external pressure increases, which is associated with the $c$-$f$ hybridization. In correspondence, the optical conductivity $\sigma(\omega)$ is also dramatically altered by pressure. The mid-infrared peak caused by hybridization gap moves to higher energies with pressure increasing, indicating the persistent enhancement of hybridization strength.

Oeschler \emph{et al} studied the low temperature thermal expansion of CeCoIn$_5$ and CeIrIn$_5$ compound under uniaxial pressures\cite{Prozorov2004}. It is found that pressures along both the axial $a$ and $c$ directions lead to an enhancement of $T_c$ in CeCoIn$_5$.
It is well known that the the highest superconducting transition temperature $T_c$ of a heavy fermion metal regularly emerges in the vicinity of the quantum critical point. The increasing of $T_c$ under pressure along both axis indicates the increasing of hybridization strength, in consistent with the optical results. As a contrast, $T_c$ of CeIrIn$_5$ increases with the application of pressure along the $a$ axis but decreases along $c$ axis, which is also evidenced by the specific heat measurement conducted by Dix \emph{et al}.\cite{Dix2009}. The positive pressure along $a$ axis and negative pressure along $c$ axis both tune the lattice constants of CeIrIn$_5$ to approach that of CeRhIn$_5$. Following this tendency, CeRhIn$_5$ is supposed to show higher $T_c$ than CeIrIn$_5$, whereas it is actually of antiferromagnetic order, due to weaker hybridization strength. This deviation demonstrated that the simply tuning of lattice constant can not compensate the huge discrepancy of $\Tilde{V}$ in different materials of the Ce$_m$M$_n$In$_{3m+2n}$ family. There must be some other factors that fundamentally modified $\Tilde{V}$.

\subsubsection{Crystal field effect  and  anisotropic hybridization}

\begin{figure}[htbp]
  \centering
  \includegraphics[width=6cm]{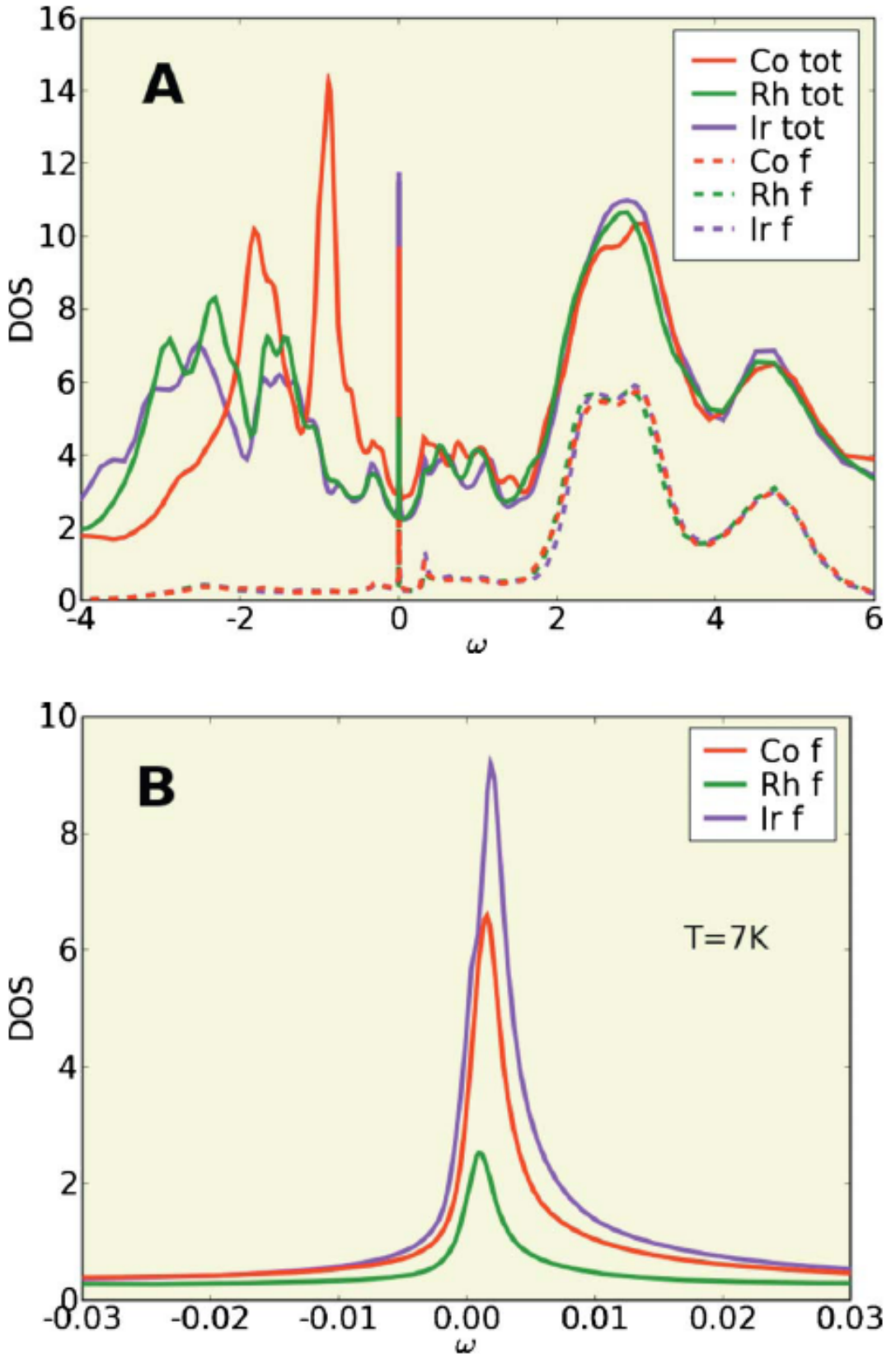}\\
  \caption{Total density of states (solid lines) and partial Ce 4$f$ density of states (dashed lines) for CeCoIn$_5$, CeRhIn$_5$, and CeIrIn$_5$ materials, in units of states per electron volt. The bottom panel shows the low-energy part of the Ce 4$f$ denstiy of states for all three compound.\cite{Haule2010}}\label{Fig:haul1}
\end{figure}

The electronic structures of the three members of CeMIn$_5$ have been calculated by Haule \emph{et al}. using the charge self-consistent combination of DFT and DMFT in the augmented plane-wave and the linear muffin-tin orbital methods\cite{Haule2010}. Actually, this method has been applied to study the optical conductivity and spectral function of CeIrIn$_5$, which agrees well with the experimental data\cite{Shim2007}. The total density of states (DOS) and the partial Ce 4$f$ DOS obtained by this approach are shown in Fig. \ref{Fig:haul1}. The major difference of the total DOS lies around 2 eV below the Fermi level, which corresponds to the contributions of transition metals. As a contrast, the partial DOS of Ce 4$f$ orbitals are quite similar for all the three compounds, except for the very low energy part, which is expanded and shown in the lower panel of Fig. \ref{Fig:haul1}.
In heavy fermion systems, the density of 4$f$ quasiparticles is a yard stick of the itineracy and hybridization. It is clearly seen that the quasiparticle peak of CeIrIn$_5$ is the largest, and CeRhIn$_5$ is the smallest. That is, CeIrIn$_5$ compound holds the strongest hybridization and is more itinerant than its two cousins.

The localization nature of CeRhIn$_5$ compound has been well demonstrated by multiple experiments, whereas the relative hybridization strength between CeCoIn$_5$ and CeIrIn$_5$ is not very clear, since their ground states are both superconductivity. The temperature dependence of nuclear spin-lattice relaxation rate of CeIrIn$_5$ measured by Nuclear quadrupolar resonance experiment suggested that the compound was much more itinerant than other Ce compounds\cite{Zheng2001a}. Therefore, the lower $T_C$ of CeIrIn$_5$ compared with CeCoIn$_5$ was possibly ascribed to the too itineracy nature of the Ir compound.
However, the measurements with application of hydrostatic pressure are not very supportive of this scenario\cite{Borth2002}.

\begin{figure}[htbp]
  \centering
  \includegraphics[width=7cm]{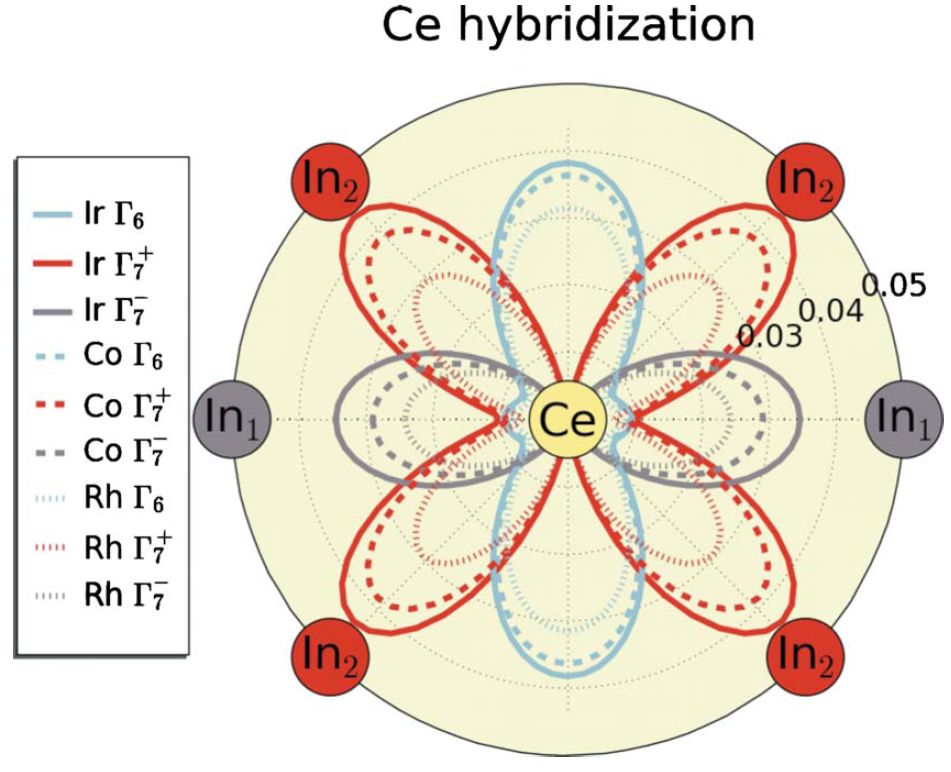}\\
  \caption{The two-dimensional projection of Ce 4$f$ hybridization functions, which were decomposed into three relevant orbitals by crystal field components of tetragonal field.\cite{Haule2010}}\label{Fig:haul2}
\end{figure}

The quasiparticle peaks are almost entirely from the band with total angular momentum $j=5/2$, the degeneracy of which is further lifted by the tetragonal crystal field environment. Therefore, the hybridization is constituted by three different components: $\Gamma_6$, $\Gamma_7^+$ and $\Gamma_7^-$. Their two dimensional projection of the corresponding hybridization function is plotted in Fig. \ref{Fig:haul2}. The $\Gamma_7^+$ orbitals point to the out of plane In$_2$ atoms while $\Gamma_7^-$ points to the in plane In$_1$ atoms. For the Ir and Co compounds, the hybridization associated with $\Gamma_7^+$ are the strongest, which is consistent with the result of uniaxial pressure experiments that the hybridization strength is most sensitive to $c$-axis distortions\cite{Prozorov2004}. The $\Gamma_6$ component is also quite strong, which points to the transition metal. In the first part of section \ref{sec:IntroCMI}, we have demonstrated that the optical conductivity of CeCoIn$_5$ and CeIrIn$_5$ shows double peak features in the mid-infrared range, which are interpreted as a consequence of the momentum dependence of hybridization. Here, it further reveals the origination of different types of hybridization.

For CeRhIn$_5$ compound, as a contrast, the $\Gamma_6$ component shows the strongest hybridization, which infers that the hybridization gap identified by optical conductivity as shown in Fig. \ref{Fig:mena2} mainly stems from the Ce 4$f$ and Rh 5$d$ electrons coupling.
We noticed that the CePt$_2$In$_7$ compound also experiences antiferromagnetic phase transition at $T_N$ = 5.2 K, even though it is much more two dimensional than all the other members in the Ce$_m$M$_n$In$_{3m+2n}$ family, which is supposed to advantage superconductivity. Remarkably, the distance between Ce and transition metal atoms in CePt$_2$In$_7$ (5 ${\AA}$) is much longer than that in CeRhIn$_5$ (3.77 ${\AA}$)\cite{Bauer2010}. Although there are no theoretical analysis on the anisotropy of hybridization in CePt$_2$In$_7$, it is very likely that the long distance between Ce and Pt substantially weakens the hybridization strength, which further leads to the AFM ground state.

It is still intriguing why the hybridization strength of CeRhIn$_5$ is the weakest. In order to resolve this question, Haule \emph{et al}. tried to calculate all the three compounds with the same lattice structure of CeIrIn$_5$. Surprisingly, the main characters remains unchanged, except for a slightly enhance of the itineracy of CeRhIn$_5$. It elaborates that the chemical difference between the three transition metals, that they hold 3$d$, 4$d$, and 5$d$ orbitals respectively, are the driving force of different levels of itineracy.

 Maehira \emph{et al} also discussed the crystal field effect for CeMIn$_5$ compounds (M=Ir and Co) based on the comparison between the relativistic band-structure calculation by RLAPW method and the dispersion obtained from a tight-binding model by including \emph{f-f} and \emph{p-p} hoppings as well as \emph{f-p} hybridization, and concluded that the crystal field effect played an important role in determining the electronic properties\cite{Maehira2003}.

\begin{figure}[htbp]
  \centering
  \includegraphics[width=7cm]{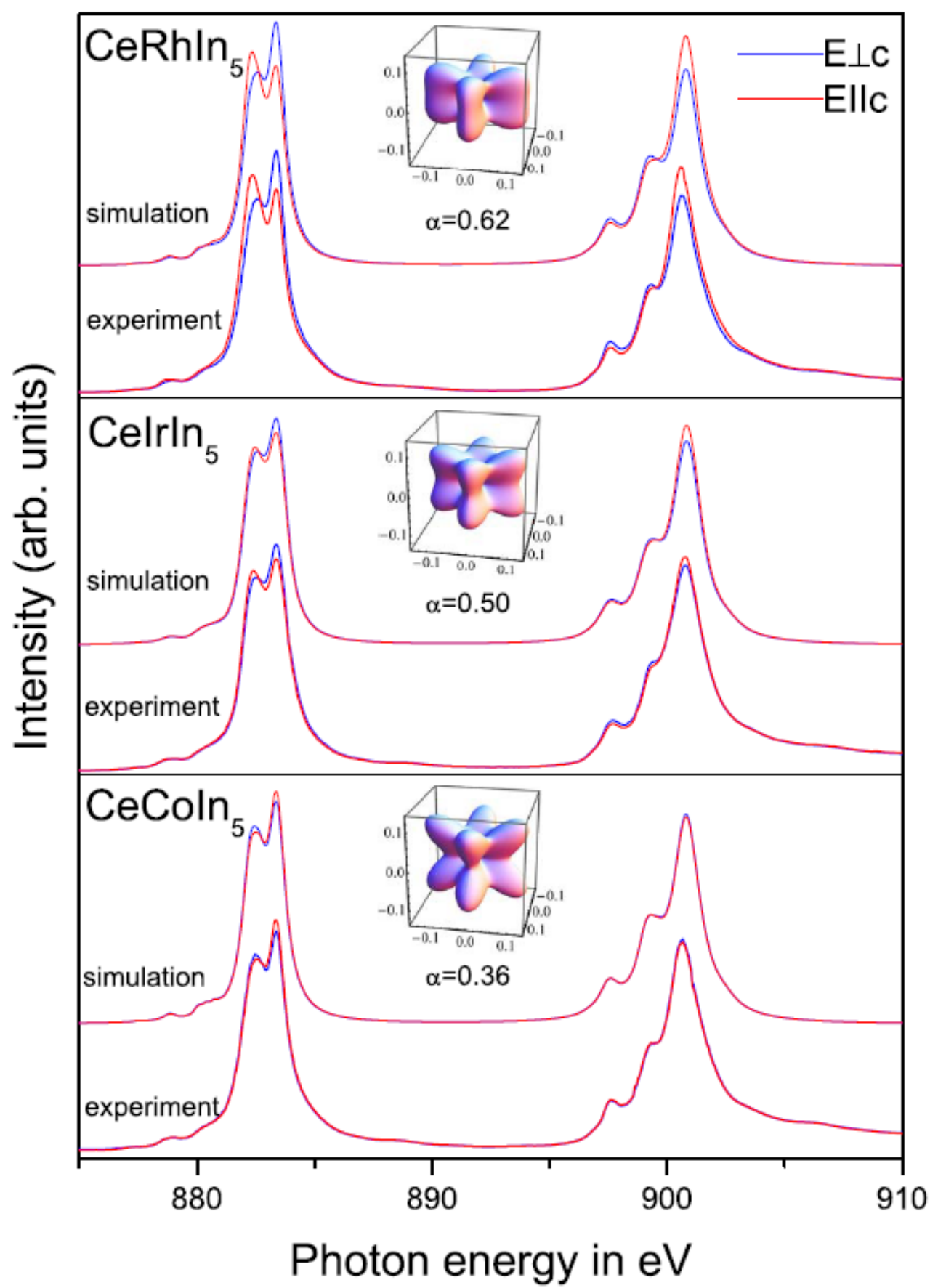}\\
  \caption{The linear-polarized XAS spectra of CeRhIN$_5$, CeIrIn$_5$ and CeCoIn$_5$ taken at 20 K. The solid lines are measured data and the dotted ones are simulations. The orbitals represent the spatial distribution of the 4$f$ wave functions according to the respective ground state admixture. \cite{Willers2010a}}\label{Fig:willers1}
\end{figure}

The importance of crystal field effect and its direct relevance to the \emph{f}-conduction electron hybridization were also addressed in many experimental studies \cite{WKnafoSRaymond2003,Christianson2004,Willers2009}. Significantly, the crystal field splitting in three 115 compounds has been measured by Willers \emph{et al.} through linear-polarized soft-x-ray absorption (XAS) and inelastic neutron scattering (INS) experiments \cite{Willers2009}. The Hund's rule ground state of Ce$^{3+}$ with $J$ =5/2 is split by tetragonal crystal into three degenerated doublet states $j_z=\pm5/2$, $\pm$3/2, and $\pm$1/2. The linear-polarized XAS spectra of CeRhIN$_5$, CeIrIn$_5$ and CeCoIn$_5$ at low temperature are shown in Fig. \ref{Fig:willers1}, which are simulated by ionic full multiplet calculations. Through this procedure, the ground state wave functions of the compounds could be obtained in the form of a mixture of $j_z=\pm5/2$ and $j_z=\pm3/2$:
\begin{equation}\label{Eq:Rh}
\begin{split}
 CeRhIn_5:\  &|0\rangle=0.62|\pm5/2\rangle+0.78|\pm3/2\rangle,\\
 CeIrIn_5:\  &|0\rangle=0.50|\pm5/2\rangle+0.87|\pm3/2\rangle,\\
 CeCoIn_5:\  &|0\rangle=0.36|\pm5/2\rangle+0.93|\pm3/2\rangle.
 \end{split}
\end{equation}
The spatial distribution of the 4$f$ wave functions are displayed in the inset of Fig. \ref{Fig:willers1}. The pure $|5/2\rangle$ orbital is actually of donut shape, while the pure $|3/2\rangle$ orbital is yo-yo shaped\cite{Willers2009}. Therefore, the wave function of 4$f$ quasiparticles in CeRhIn$_5$ holds the most flat spatial distribution, since there are more amount of $j_z=\pm5/2$ components. It has been demonstrated that the hybridization in CeMIn$_5$ compounds is dominated by the interaction between Ce and out of plane In atoms. A flat 4$f$ orbital is an apparently disadvantage of this kind of hybridization. The flatness follows a general sequence Rh $>$ Ir $>$ Co. Assuming the difference in hybridization strength $\Tilde{V}$ is mainly caused by CFE,   $\Tilde{V}$ would follow the trend Rh $<$ Ir $<$ Co. This indeed makes sense, as the ground state of Rh is antiferromagnetic, Ir superconductivity with $T_C$ =0.4 K, and Co with $T_C$ = 2.3 K. Although the DMFT calculation does not agree with this conclusion, it still derived some implications on the exotic order of the hybridization strength in the CeMIn$_5$ family.

\section{Infrared properties of mixed-valent systems}

Mixed-valent or strong valence fluctuation compounds with rare earth or actnide elements both refer to the materials whose atoms possess a non-integer valence, such as YbAl$_3$, EuRh$_2$, TmSe, and YbAl$_2$ etc. Distinctive with the charge order materials in which the atoms of the same element locating at inequivalent sites exhibit different valences, the mixed-valent compound holds homogeneous charge configurations. Taking lanthanides as examples, the low temperature ambivalent behavior is ascribed to the nearly degeneracy of 4$f^n$(5$d$6$s$)$^m$ and 4$f^{n-1}$(5$d$6$s$)$^{m+1}$ states. At high temperatures well above $T_K$, the distribution of $f$ electrons are stable obeying the Hunds rule, giving rise to an integer valence. However, when the temperature decrease, the 4$f$ electrons hybridize with conduction electrons and open up a gap at the Fermi surface. The hybridization enables the electron migrations associated with the two degenerated states and the valence of lanthanides fluctuates between 4$f^n$ and 4$f^{n-1}$ as a consequence.

Mixed valence systems are commonly observed in the intermetallic compounds with Ce, Yb, Sm, Eu, U and other rare earth elements. In particular,Yb compounds attract a great deal of interest because the trivalent Yb ion is in some sense the hole
counterpart of the Ce$^{3+}$ ion which has one electron in its 4$f$ shell. It is worth noting that mixed-valent compounds generally have stronger hybridization strength between 4$f$ and conduction electrons than the heavy fermion paramagnetic metals or antiferromagnets due to reduced lattice constants, therefore the hybridization-induced crossover of 4$f$ electrons
from high-temperature local moment state to low-temperature Fermi-liquid state usually appears at higher temperature scale. An associated enhanced Pauli paramagnetism is also observed over wide temperature range. Here we review infrared properties of several mixed-valent compounds with different hybridization strength with an emphasis on the scaling between the hybridization energy gap and the Kondo temperature or quantity related to the hybridization strength.

\subsection{YbAl$_3$}

YbAl$_3$ is a well known mixed valent compound. The mean valence of Yb is about 2.65 - 2.8\cite{Okamura2004}. Its specific heat coefficient is $\gamma\sim$ 40 mJ/(mol$\cdot$K$^2$). There are several temperature scales for the compound: the Kondo temperature $T_K$ is about 670 K, the coherence temperature $T_{coh}$ is about 35 K, and the $T_{max}$ (\emph{i.e.} the temperature where magnetic susceptibility $\chi(T)$ shows a peak) is about 120 K\cite{Cornelius2002}. Usually investigations of YbAl$_3$ have been accompanied by those of LuAl$_3$ which can be used as a reference material because its 4$f$ states are fully occupied and located well below the Fermi level \cite{Okamura2004}.

\begin{figure}
  \centering
  \includegraphics[width=8cm]{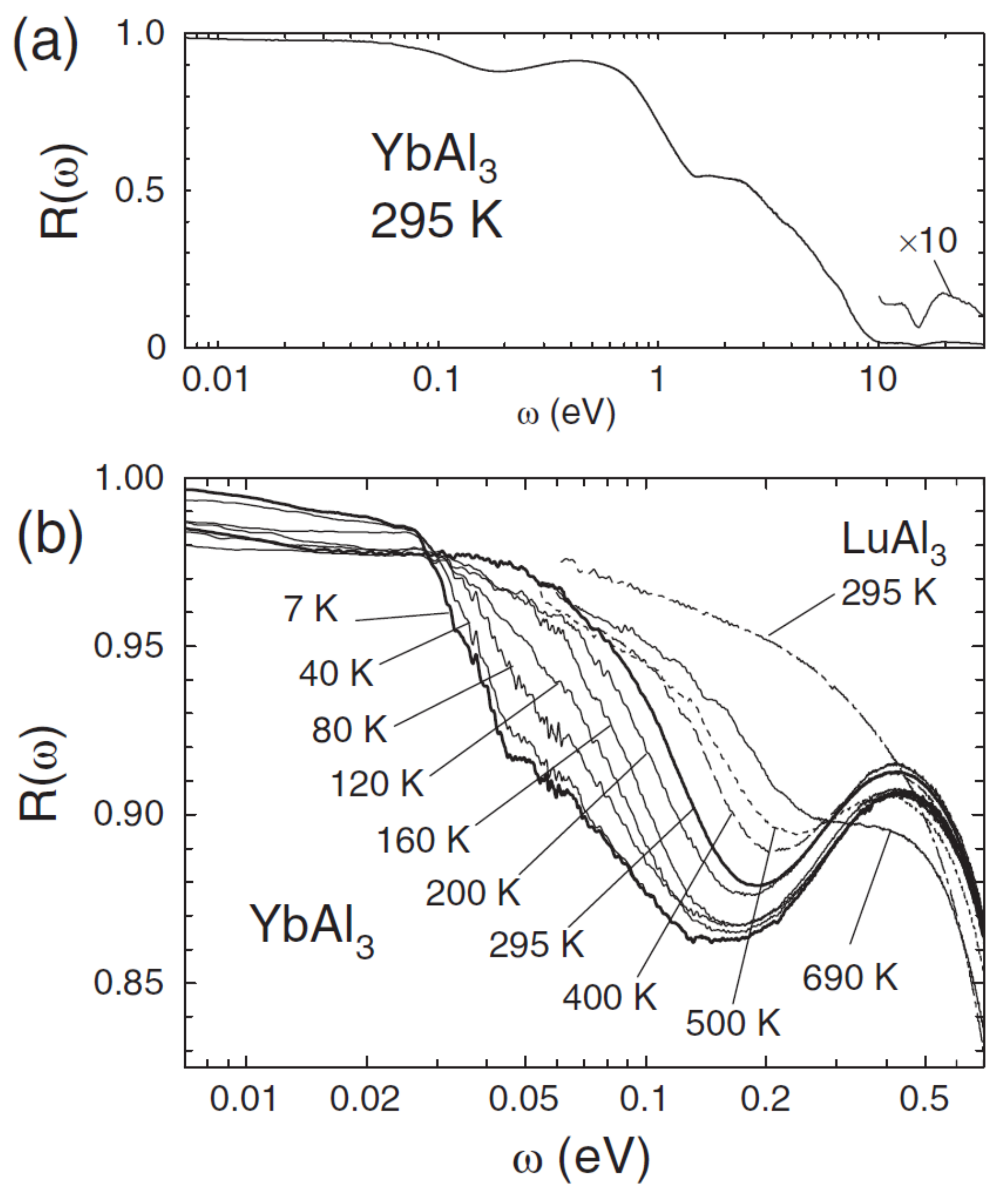}\\
  \caption{(a) Optical reflectivity spectrum of YbAl$_3$ at room temperature over broad energies. (b) $R(\omega)$ of YbAl$_3$ between 8 and 690 K, and that of LuAl$_3$ at 295 K. \cite{Okamura2004}}\label{Fig:YbAl3-1}
\end{figure}

The optical measurement was reported by Okamura et al.\cite{Okamura2004a,Okamura2004}. Besides the generic features commonly seen for other heavy fermions and mixed valent compounds, like the narrow Drude component and the mid-infrared peak, the optical conductivity shows peculiar structure below the mid-infrared peak, which stimulated some further discussions. Figure \ref{Fig:YbAl3-1} shows reflectivity $R(\omega)$ spectra of YbAl$_3$. $R(\omega)$ is suppressed below 0.4 eV leading to a broad dip, which becomes more pronounced with decreasing T. In contrast, LuAl$_3$ has no such feature in $R(\omega)$. These results
indicate that the dip in $R(\omega)$ of YbAl$_3$ is caused by the Yb 4$f$-related electronic states located near the Fermi level.

\begin{figure}
  \centering
  \includegraphics[width=8cm]{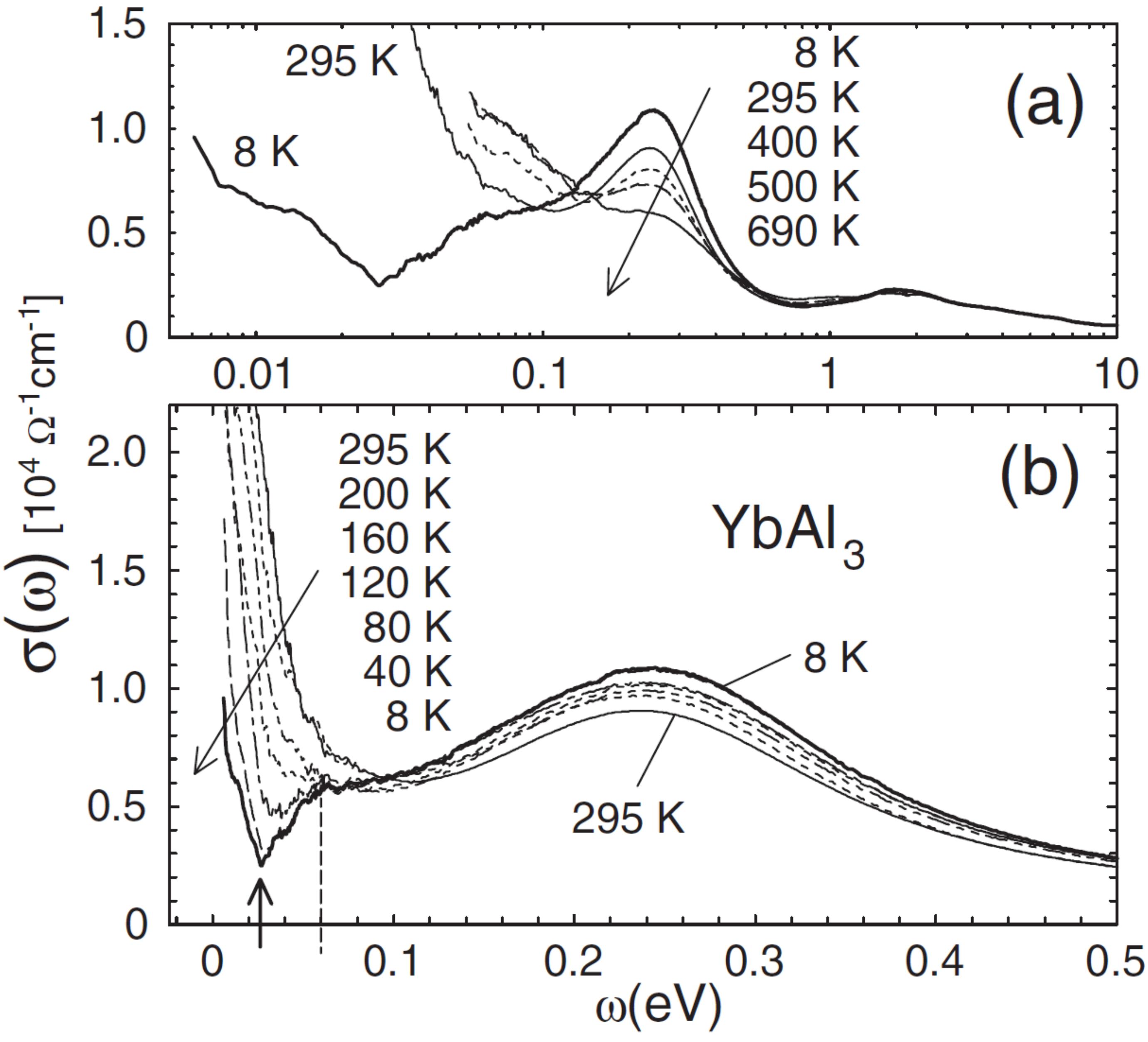}\\
  \caption{(a) and (b) Optical conductivity $\sigma_1(\omega)$ spectra of YbAl$_3$. In (b) the
vertical broken line and the vertical arrow indicate the shoulder and the pseudogap, respectively, developed at temperatures below 120 K.\cite{Okamura2004}}\label{Fig:YbAl3-2}
\end{figure}

Figure \ref{Fig:YbAl3-2} shows the real part of the optical conductivity of YbAl$_3$ at different temperatures. The conductivity $\sigma_1(\omega)$ shows a narrow Drude component at very low frequency and a pronounced mid-infrared peak near 0.25 eV. Those are generic features for HF or mixed valent systems. However, what is peculiar is the appearance of a strong depletion (which they refers to as a "pseudogap") of $\sigma_1(\omega)$ below 120 K, indicated by the vertical arrow in Fig. \ref{Fig:YbAl3-2}(b), and the associated shoulder at 60 meV, indicated by the vertical broken line. They emphasized that the pseudogap in $\sigma_1(\omega)$ is not merely a tail of the mid-IR peak but is a distinct feature, since the former appears below 120K ($\sim T_{max}$) and becomes well developed only at 40K and below, while the latter is observed up to much higher temperatures.

Since the shoulder and the pseudogap are not observed in all HF and mixed-valent compounds, Okamura et al. speculated that it may not be explained on the basis of simple PAM. Instead, they interpreted the mid-IR peak as arising from the transition across the direct hybridization energy gap $\Delta_{dir}$, while the shoulder or pseudogap is ascribed to the indirect transition across the hybridization gap $\Delta_{ind}$. They argued that the energy scale of the pseudogap $\sim$60 meV is close to the Kondo temperature $T_K$ of YbAl$_3$. \cite{Okamura2004}. On the other hand, Vidhyadhiraja and Logan \cite{Logan2005,Vidhyadhiraja2005} employed a local moment approach to the PAM within the framework of dynamical mean-field theory and found that a proper choice of material parameters for the PAM could reproduce the shoulder feature at 50$\sim$60 meV followed by a direct hybridization gap peak at 0.25 eV. So the shoulder or pseudogap also has a origin related to the correlation effect of f-electron self-energy.

It appears to us that the shoulder feature observed in YbAl$_3$ is similar to that seen in CeCoIn$_5$ below the mid-IR peak. It could be caused by the presence of multiple conduction bands of the material which cross the Fermi level and hybridize with the localized 4$f$ band or originated from the momentum-dependent hybridizations of conduction bands to the 4$f$ electrons.

Optical studies on related Yb-based mixed valent compounds, e.g. YbRh$_2$Si$_2$ and YbIr$_2$Si$_2$, were also reported \cite{Kimura2006,Iizuka2010a}. YbRh$_2$Si$_2$ locates very close to an antiferromagnetic QCP in the phase diagram where the valence state of Yb is close to 3, while YbIr$_2$Si$_2$ locates within the mixed valent regime with smaller valence fluctuations than that of YbAl$_3$ \cite{Guritanu2012}. Those compounds show similar hybridization energy gap features but with slightly lower energy scales, in accord with relatively weak hybridization strength. What is different is that they exhibit non-Fermi-liquid behaviors in the coherent heavy qusiparticle state. For example, YbRh$_2$Si$_2$ exhibits a linear frequency dependence of the low-energy scattering rate\cite{Kimura2006}, while YbIr$_2$Si$_2$ shows a power-law frequency  dependence\cite{Iizuka2010a}, similar to the linear or power law dependence of temperature found in the electrical resistivity measurement on those compounds.  It may be worthwhile to mention that the linear temperature or frequency dependence often found near QCPs could be simply a consequence of the low electronic scale that gives rise to the scattering following classical limit. This was first noted theoretically by Kaiser and Doniach in 1970\cite{Kaiser1970}.

\subsection{YbIn$_{1-x}$Ag$_x$Cu$_4$ and scaling between $\Delta_{dir}$ and $\sqrt{T_K}$}

YbInCu$_4$ shows valence transition at about 40 K which is accompanied with abrupt changes in the
physical quantities such as lattice parameters, electrical resistivity, magnetic susceptibility $\chi(T )$ and nuclear
spin-lattice relaxation rate 1/$T_1$ \cite{Cornelius1997,Kishimoto2003}. Above this temperature $\chi(T )$ obeys the Curie-Weiss law with the effective moment close to that of Yb$^{3+}$ ion, while below the valence transition temperature, YbInCu$_4$ shows Pauli paramagnetism, and its average valence
is 2.9$+$. With Ag substitutes for In, the Kondo temperature could be effectively tuned.

\begin{figure}
  \centering
  \includegraphics[width=6cm]{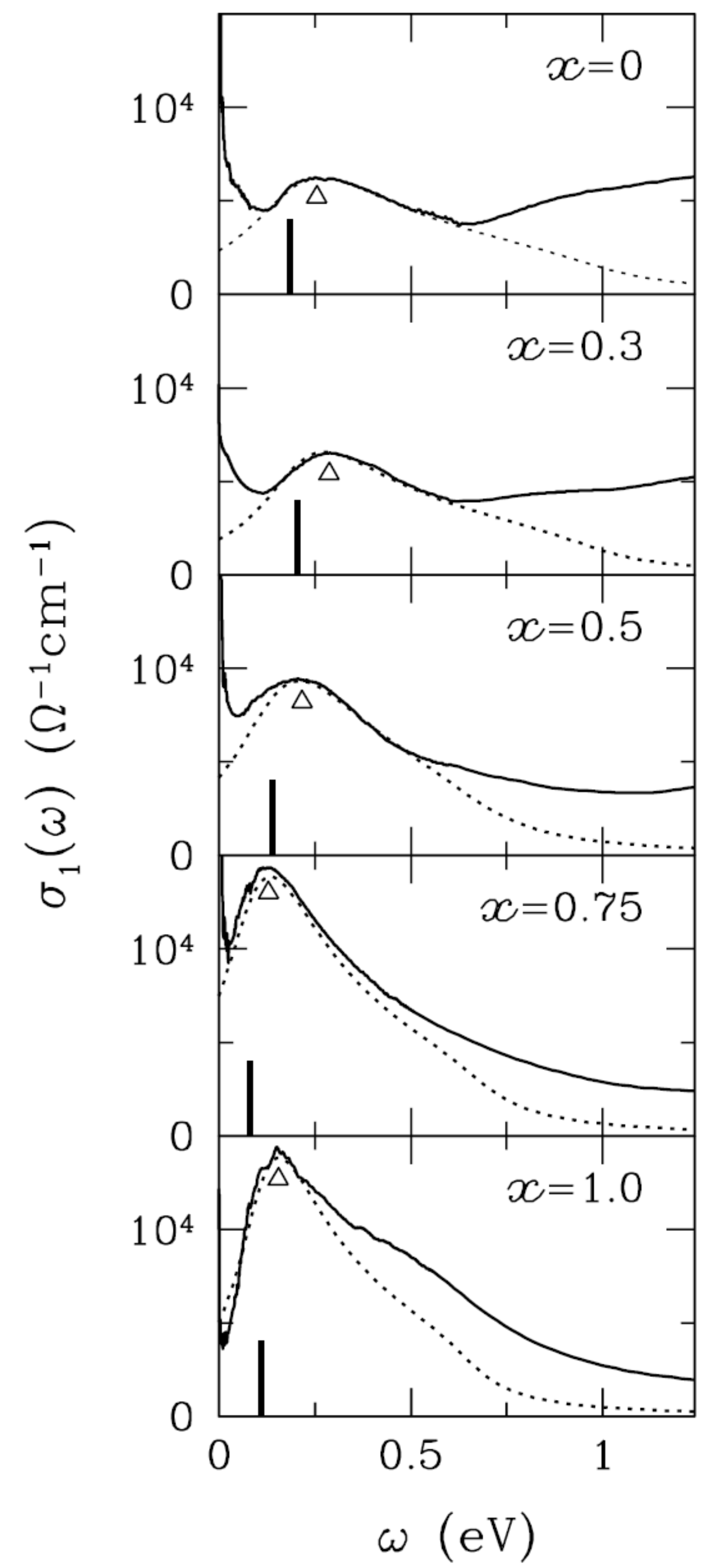}\\
  \caption{The real part of optical conductivity of YbIn$_{1-x}$Ag$_x$Cu$_4$ for five different x measured at 20 K. The open triangles mark peak frequencies. The dotted curves refer to a mean-field approximation calculation to the PAM. The dark vertical bars indicate a threshold frequency 2$\tilde{V}$ from PAM model.\cite{Hancock2004}}\label{Fig:YbInCu-1}
\end{figure}

Garner et al. studied the optical response of  YbInCu$_4$ first and pointed out that the position of mid-infrared peak in conductivity spectrum should obey Equ.\ref{PAM3} in PAM \cite{Garner2000a}. Later, Hancock et al. studied the optical properties of Ag-doped YbInCu$_4$ and found a scaling behavior between the hybridization energy gap and the Kondo temperature \cite{Hancock2004}. Figure \ref{Fig:YbInCu-1} shows the real part of the optical conductivity of YbIn$_{1-x}$Ag$_x$Cu$_4$ at 20 K for five values of x. Similar to HF systems, the conductivity shows a narrow Drude component at very low frequency, a suppression dip and broad mid-infrared peak. The peak structure is associated with the development of the hybridization energy gap, which is not present at high temperature\cite{Garner2000}. Hancock et al. found that the peak could be reproduced by the PAM (Equ. \ref{PAM3}) with a proper choosing of parameters in the model together with a broadening factor $\Delta$ \cite{Hancock2004}. For example, for x=0.75 compound, the parameters are $\tilde{V}$=41 meV, $\tilde{\epsilon}_f $=2.5 meV, $E_F$=1eV, $\mid\textbf{P}_{+,-}\mid^2k_F$=4.673${\AA}^3$, and broadening factor $\Delta$=0.12 eV. The calculated data are presented as dash curves in Fig. \ref{Fig:YbInCu-1}. According to the calculations, the threshold of the optical transition $\Delta_{dir}$=2$\tilde{V}$ (marked as vertical bar in the figure) is somewhat different from the peak position (marked as open triangle). The mid-infrared peak, which centers around 0.25 eV for x=0 (YbInCu$_4$), is found to change only slightly as x is increased to 0.3; however, further doping to x=0.5-0.75 causes both a redshift and strengthening of this feature.

\begin{figure}
  \centering
  \includegraphics[width=8cm]{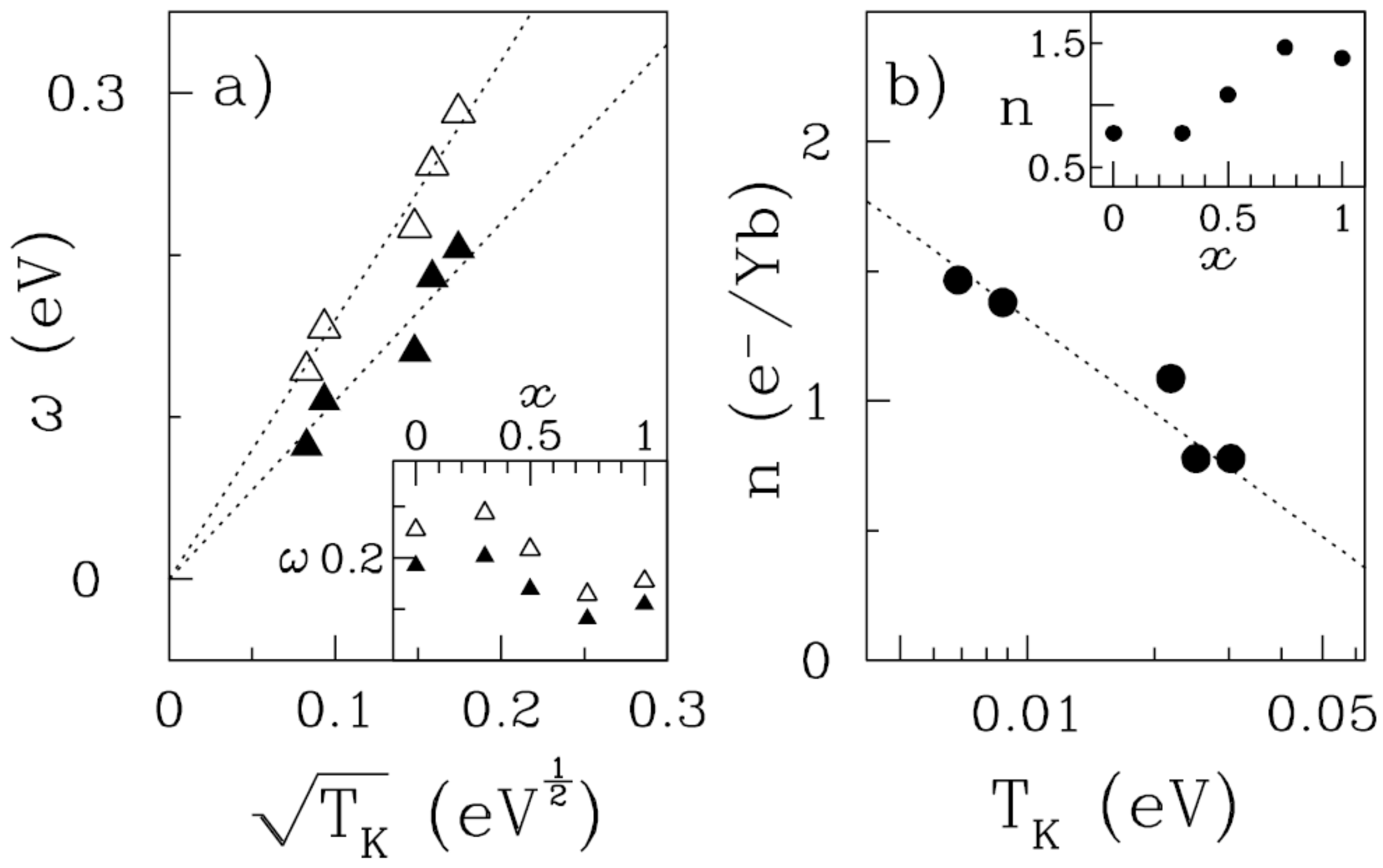}\\
  \caption{ (a) Characteristic frequencies of peak ($\triangle$) and threshold ($\blacktriangle$) as a function of x (inset) and the square-root of Kondo temperature $\sqrt{T_K}$ for YbIn$_{1-x}$Ag$_x$Cu$_4$. (b) Spectral weight of mid-infrared peak as a function of x (inset) and the Kondo temperature $T_K$.\cite{Hancock2004} }\label{Fig:YbInCu-2}
\end{figure}

The Kondo temperature $T_K$ is known to change with the Ag concentration\cite{Cornelius1997}. Hancock et al. plotted the two characteristic frequencies, corresponding respectively to the peak and threshold of this excitation, both as a function of $\sqrt{T_K}$ showing in the main panel and x in the inset of Fig. \ref{Fig:YbInCu-2}. Values of $T_K$ as a function of x are obtained from measurements of magnetic susceptibility \cite{Cornelius1997}. They indeed found a good scaling between the hybridization energy gap and the Kondo temperature, $\Delta_{dir}\propto \sqrt{T_K}$, in accord with the prediction by the PAM, i.e. Equ.\ref{PAM3} in the Introduction section. They also found that the strength of the peak could also be explained by the mean-field approximation of PAM.  The variation of the spectral features in YbIn$_{1-x}$Ag$_x$Cu$_4$ provides a very crucial test of the correlations established in PAM since one may expect that only the hybridization and valence may change significantly through the transition, unlike comparison between different materials.

\subsection{Universal scaling in Ce- and Yb-based compounds}

From optical studies on Ce$_m$M$_n$In$_{3m+2n}$ family and above-discussed Yb-based compounds, it is clear that the hybridization energy gap determined by the optical spectroscopy is correlated with the hybridization strength. A scaling between the hybridization energy gap and the square root of Kondo temperature predicted by the PAM is found for YbIn$_{1-x}$Ag$_x$Cu$_4$ series \cite{Hancock2004}. Okamura et al. examined more Ce- and Yb-based compounds and found that a better scaling could be achieved if the bandwidth of conduction electrons \emph{W} was taken into account, that is $\Delta_{dir}\propto \sqrt{WT_K}$\cite{Okamura2007}.

\begin{figure}
  \centering
\includegraphics[width=8cm]{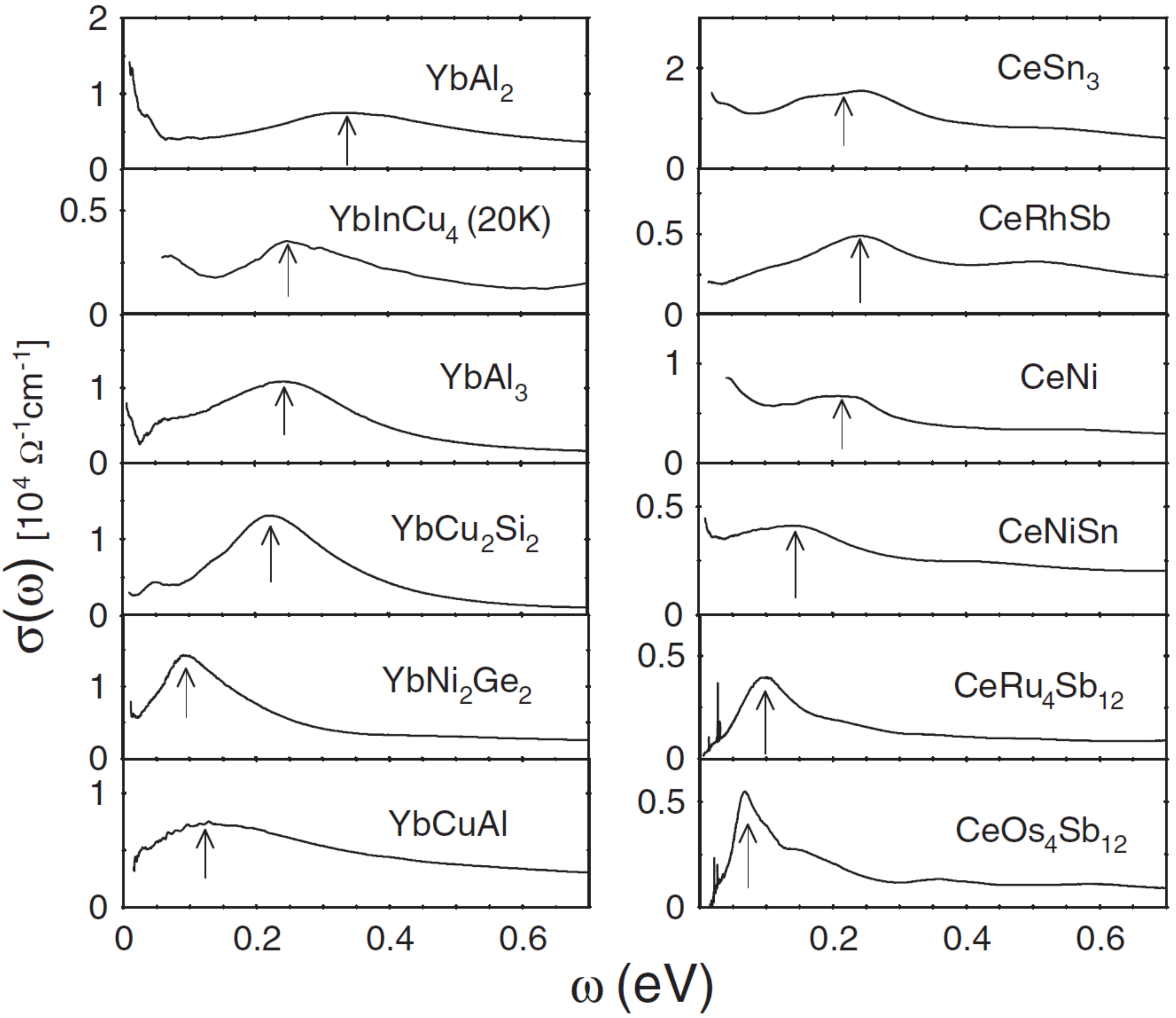}\\
  \caption{$\sigma_1(\omega)$ spectra of a number of Ce- and Yb-based compounds measured at 9K, except for YbInCu$_4$ measured at 20 K. The arrows indicate the positions of the mid-infrared peaks. \cite{Okamura2007}}\label{Fig:Scaling1}
\end{figure}

Figure \ref{Fig:Scaling1} presents the optical conductivity spectra $\sigma_1(\omega)$ of a number of Ce- and Yb-based compounds measured at low temperature. The arrows indicate the positions of the mid-infrared peaks. In order to see the scaling, Okamura et al. expressed the conduction electron bandwidth \emph{W} and Kondo temperature $T_K$ in terms of experimentally observable quantities. They suggested that the bandwidth \emph{W} of a Ce (Yb) compound could be regarded as inversely proportional to specific heat coefficient $\gamma$ of the non-magnetic, isostructural La (Lu) compound (denoted as $\gamma_0$). $T_K$ is expressed in terms of the electronic specific heat coefficient $\gamma$ as\cite{Hewson1993,Okamura2007}
\begin{equation}
k_BT_K={{e^{1+C-3/2N}}\over{2\pi\Gamma(1+1/N)}}{{\pi^2}\over{3\gamma}}{{(N-1)}\over{N}}
={{\pi^2}\over{3}}{{(N-1)\omega_N}\over{N}}/\gamma={{\pi^2}\over{3}}a/\gamma,
\label{scaling1}
\end{equation}
where C is the Euler's constant which has the value 0.577216, N=2j+1 is the \emph{f} level degeneracy, $\Gamma(1+1/N)$ is the $\Gamma$ function of 1+1/N, $\omega_N=e^{1+C-3/2N}/{2\pi\Gamma(1+1/N)}$ is the Wilson number. The parameter $a=(N-1)\omega_N/N$ is a constant which depends
only on the \emph{f} level degeneracy. It is easy to check that \emph{a} has the values of 0:21, 0.54, and 0.59 for N=2, 6, and 8, respectively. Then, Equ.\ref{PAM3} in PAM can be rewritten as
\begin{equation}
\Delta_{dir} \propto \sqrt{a/(\gamma\gamma_0)}.
\label{scaling2}
\end{equation}

\begin{figure}
  \centering
\includegraphics[width=8cm]{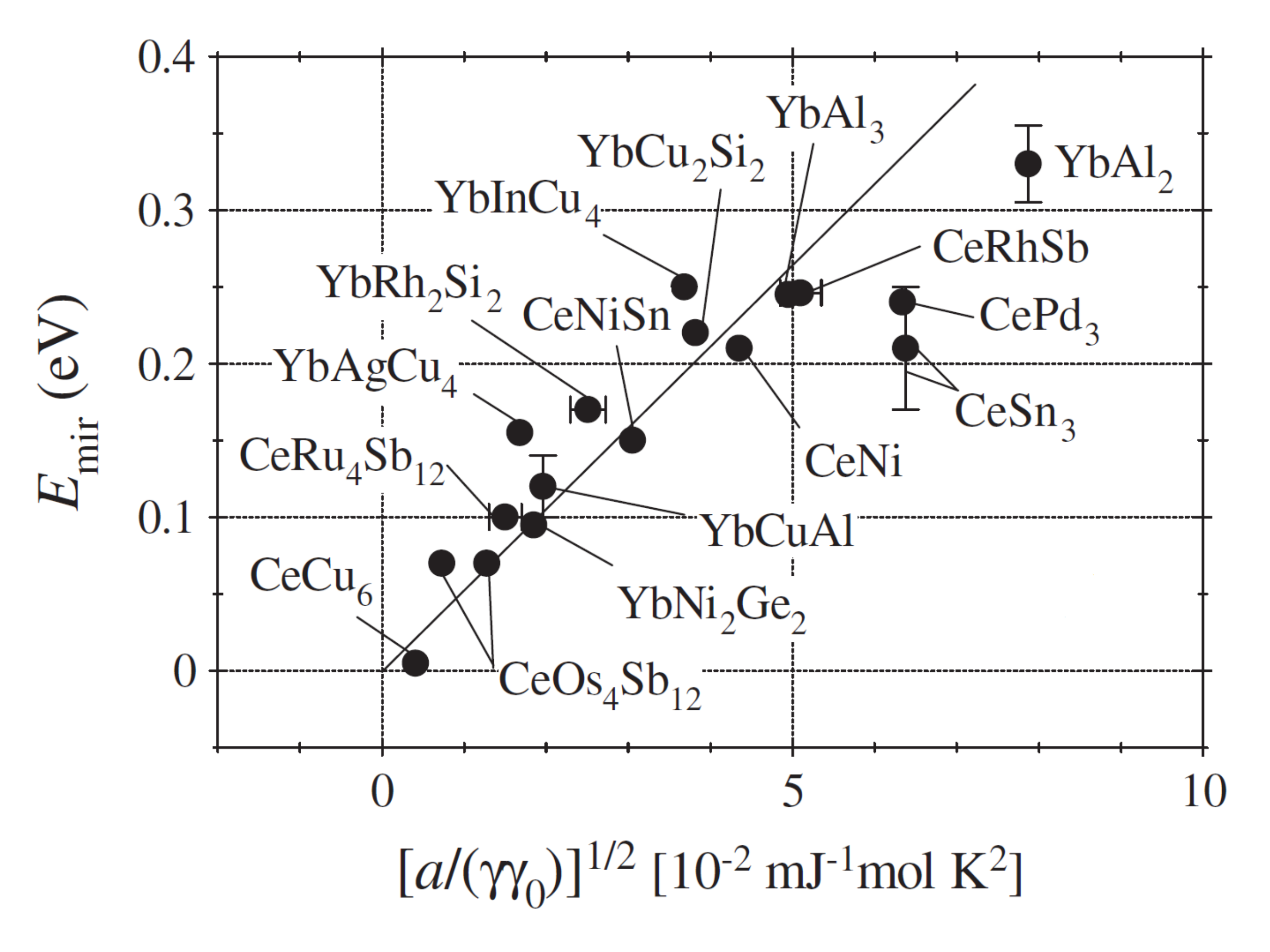}\\
  \caption{Mid-infrared peak energy plotted as a function of $\sqrt{a/(\gamma\gamma_0)}$ for many heavy fermion compounds. The solid line is a guide to the eyes.\cite{Okamura2007}}\label{Fig:scaling2}
\end{figure}

By taking the specific heat coefficients ($\gamma$) of the Yb (Ce) compounds and those ($\gamma_0$) of the isostructural Lu (La) compounds reported in literature and taking account of proper degeneracy of \emph{f}-level, they plotted the mid-infrared peak energies (which is close to the $\Delta_{dir}$) as a function of $\sqrt{a/(\gamma\gamma_0)}$ for a number of HF and mixed valent compounds and indeed found a good linear dependent behavior. They also demonstrated that similar plot without including $\gamma_0$ ($\propto W^{-1}$) is less satisfactory in its linear dependence \cite{Okamura2007}. The relatively good scaling seen in previous study\cite{Hancock2004} on YbIn$_{1-x}$Ag$_x$Cu$_4$ without including bandwidth could be due to the fact that the bandwidth of conduction electrons actually changes very small in the series. The good scaling between hybridization energy gap $\Delta_{dir}$ and the hybridization strength $\sqrt{WT_K}$ demonstrates that the PAM can generally and quantitatively describe the low-energy charge excitations in a wide range of HF compounds. It is speculated that this scaling is universal for all Ce- and Yb-based compounds.

\subsection{Other mixed valent compounds}

Optical investigations on heavy fermions and mixed valent compounds have been performed mostly on Ce-, U- and Yb-based compounds. Yet there are some reports on other rare-earth based systems, for example, on Sm- and Eu-based compounds. Here, as examples, we discuss optical spectra on two Eu-based compounds EuIr$_2$Si$_2$ and EuNi$_2$P$_2$ \cite{Guritanu2012}. Usually, Eu-based systems show somewhat different properties as a function of pressure \emph{P} or composition \emph{x} as compared to Ce- or Yb-based systems. In Ce- or Yb-based mixed valent compounds, increasing the hybridization usually leads to a smooth and continuous evolution of valence state as a function of \emph{P} or \emph{x}. In contrast, Eu-based valence fluctuated systems exhibit a first order valence state transition as a function of \emph{P} or \emph{x}. The difference leads to the suggestion that the the valence fluctuations in Eu-based systems have a different origin from that in Ce- or Yb-based systems \cite{Guritanu2012}. Nevertheless, the optical spectroscopy measurements on Eu-based compounds reveal surprisingly similar features at low temperature, including the presence of narrow Drude response, the partial suppression of the optical conductivity and a well-defined mid-IR peak.

\begin{figure}
  \centering
  \includegraphics[width=8.5cm]{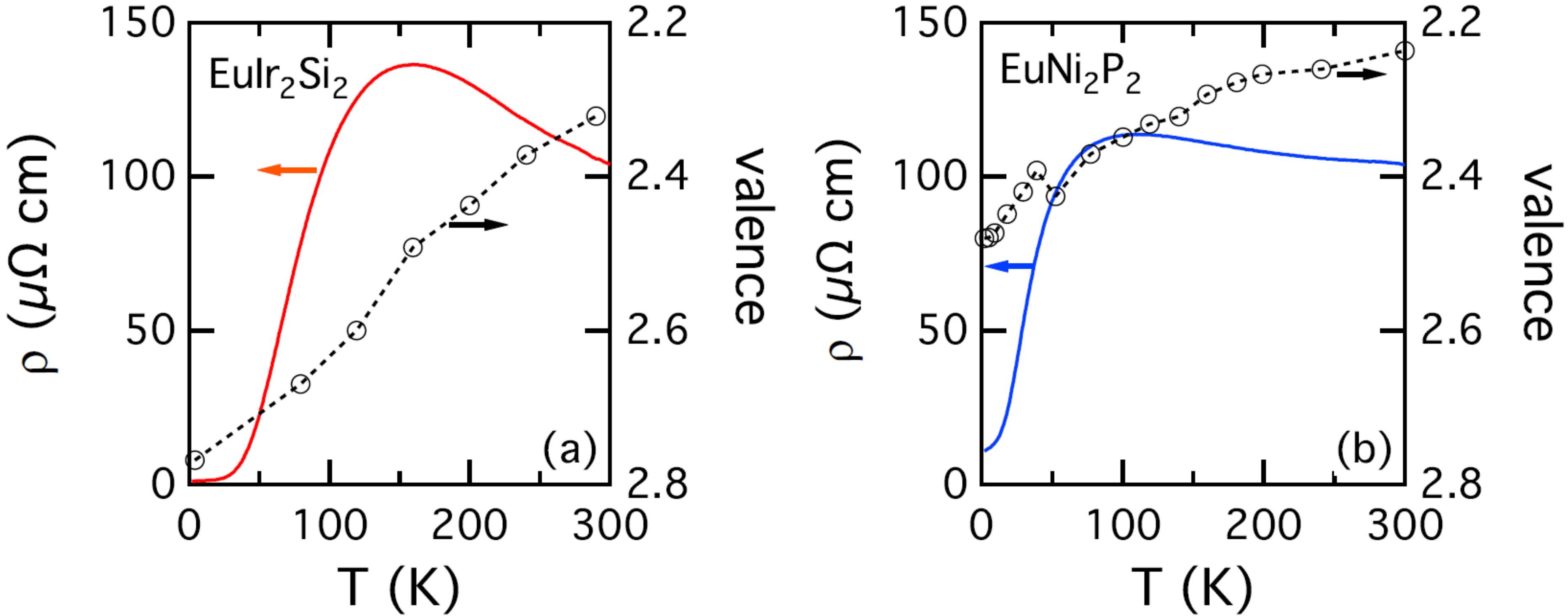}\\
  \caption{Resistivity and valence change as a function of temperature for EuIr$_2$Si$_2$ and EuNi$_2$P$_2$ \cite{Guritanu2012}. }\label{Eu122-1}
\end{figure}

EuIr$_2$Si$_2$ is a valence fluctuating Eu-system where the valence decreases from 2.8 at 4K to 2.3 at 300K. The valence state of EuNi$_2$P$_2$ at 300 K is similar to that of EuIr$_2$Si$_2$, but remains close to 2.5 at low temperature. Similar to other heavy fermion systems, their resistivity $\rho(T )$ increases upon cooling below 300 K, reaches to a maximum near 150K (100K) in EuIr$_2$Si$_2$ (EuNi$_2$P$_2$), then decreases dramatically at lower temperatures. Their valence and resistivity change as a function of temperature is shown in Fig. \ref{Eu122-1}. \cite{Guritanu2012}

\begin{figure}
  \centering
  \includegraphics[width=8cm]{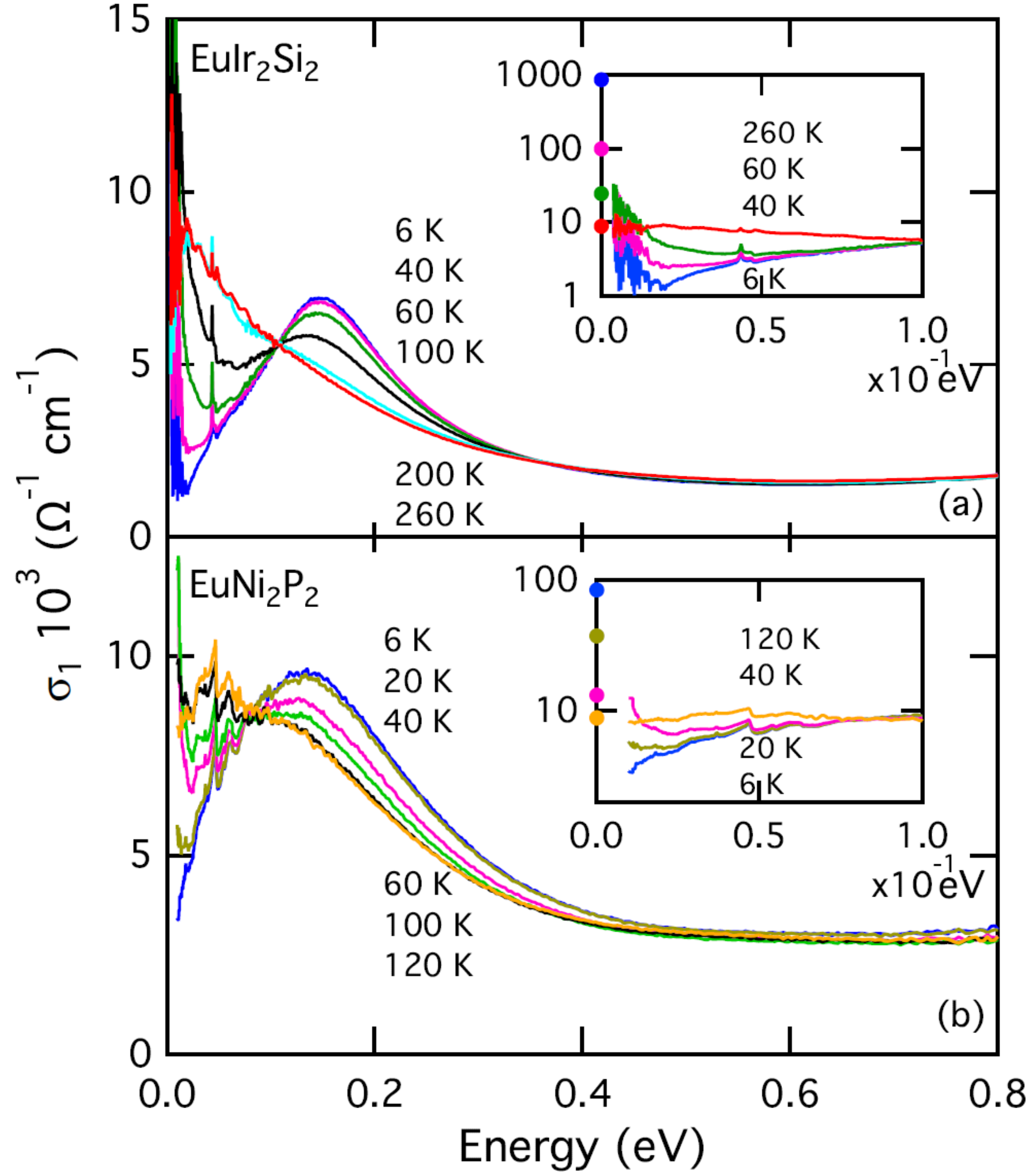}\\
  \caption{The real part of the optical conductivity spectra of EuIr$_2$Si$_2$ (a) and EuNi$_2$P$_2$ (b) at various temperatures. Insets: The low-energy optical conductivity spectra (solid lines) and the corresponding values for the dc conductivity for EuIr$_2$Si$_2$ and EuNi$_2$P$_2$. \cite{Guritanu2012} }\label{Eu122-2}
\end{figure}

Fig. \ref{Eu122-2} (a) and (b) show the real part of the optical conductivity spectra $\sigma_1(\omega)$ for EuIr$_2$Si$_2$ and EuNi$_2$P$_2$, respectively \cite{Guritanu2012}. As the temperature is lowered, three features generic to heavy fermion systems are clearly observed in both materials: the formation of narrow Drude peak arising from the renormalized heavy quasiparticles, the strong suppression of the optical conductivity below 0.15 eV reflecting the opening of hybridization energy gap and the appearance of well-defined mid-IR peaks. The peak centers at 0.15 eV for EuIr$_2$Si$_2$ and at 0.13 eV for EuNi$_2$P$_2$. The data represent the first observation of mid-IR peak in mixed valent Eu-based compounds. Despite the fact that the Eu-based systems show a different evolution of valence state as a function of pressure or chemical tuning from Ce- or Yb-based mixed valent systems, the striking similarities in optical spectra also highlight the importance of hybridization between 4$f$- and conduction electrons in the Eu-based systems.

Before concluding, we would like to point out a striking difference between Ce- (or Yb-) based HF compounds and the U-based compounds. Actually, there have been many reports of optical studies on U-based HF compounds having $f^2$ or $f^3$ configurations \cite{Degiorgi1999,Degiorgi2001}, the overall features are very similar, such as the appearance of narrow Drude component and the formation of hybridization energy gap or mid-infrared peak at low temperature. However, the mid-infrared peak resulting from the transfer of spectral weigh from low frequency to high frequency associated with the hybridization gap formation could appear at much higher energy (see figures in ref.\cite{Degiorgi2001}). For a well-known U compound, URu$_2$Si$_2$, the suppressed spectral weight in the gapped region is actually transferred to the energy centered at 3500 cm$^{-1}$ ($\sim$0.43 eV) \cite{Guo2012}, which is much higher than the mid-infrared peak energy discussed above. As it is already argued by Okamura et al.\cite{Okamura2007}, the scaling property in the hybridization band model discussed above is based on the $f^1$ configuration. It cannot directly apply to the U-based or other lanthanide HF compounds with different f-electrons configurations. To quantitatively understand the discrepancies between those compounds, further progress is required in the theoretical understanding of their electronic structures.

\section{Summary}

In summary, the behavior of strongly correlated heavy-electron systems with 4$f$ or 5$f$ orbitals are governed by the subtle competition between Kondo effect and RKKY interactions, both of which are directly related to the coupling strength $J$. When $J$ is very small, the RKKY exchange interaction plays a leading role, whereas Kondo effect dominates for large $J$ values. Consequently, as the coupling strength increases, the heavy electron materials transform from antiferromagnets to paramagnetic metals, then to mixed valent compounds.
The electromagnetic response in dense Kondo lattice could be well described by PAM, which predicts that the hybridization between conduction band and $f$-level gives rise to a gap near the Fermi level. The energy scale of direct gap $\Delta_{dir}$ is determined by the hybridization strength which is proportional to the square root of Kondo temperature $T_K$ times the conduction electron bandwidth \emph{W}. In addition, the effective mass of heavy quasiparticles is controlled by the ratio of $\Delta_{dir}/T_K$.

At high temperatures well above $T_K$, the optical conductivity of a heavy fermion or mixed valence compound generally shows a broad metallic Drude peak, connected with the strong scattering of local moments. As the temperature decreases and the Kondo coherence sets in, the Drude peak gets much sharper due to the Kondo screening of local moments and the spectral weight is substantially suppressed due to the effective mass enhancement of quasiparticles. At the same time, the collective excitations across the hybridization gap will lead to the development of a peak in the optical conductivity spectrum.

We reviewed the optical response of a number of materials with varying degrees of coupling strength, spanning from heavy fermion antiferromagnets to mixed valence compounds. The optical conductivities of all the materials exhibit emerging hybridization peaks and strong narrowing of Drude components below $T_K$, qualitatively obeying the PAM predictions. Especially, almost all the Ce- and Yb- based heavy fermion compounds can be quantitatively scaled by a universal law $\Delta_{dir}\propto \sqrt{WT_K}$, revealing the underlying relationship between the direct gap and hybridization strength. However, Eu- and other lanthanides- or actinides-based compounds with more than one $f$ electrons do not follow this scaling law, since the discussed model is based on the $f^1$ configuration.

\begin{center}
\small{\textbf{ACKNOWLEDGMENTS}}
\end{center}

We acknowledge helpful discussions with Y. F. Yang, J. Thompson, H. Q. Yuan, Q. Si and F. Steglich. This work was supported by the National Science Foundation of China (11120101003, 11327806), and the 973 project of the Ministry of Science and Technology of China (2012CB821403).

\bibliography{IR-HFv2}

\end{document}